\documentclass[aps,plb,preprintnumbers,nofootinbib]{revtex4}

\usepackage{amsmath,amsfonts,amssymb,bm}
\usepackage{graphicx}
\usepackage{color}
\usepackage{subfigure}
\usepackage{multirow}
\usepackage{textcomp}
\usepackage{slashed}
\usepackage{enumitem}
\usepackage{rotating}

\definecolor{purple}{rgb}{0.5,0,0.5}
\definecolor{blue}{rgb}{0.0,0,0.9}

\usepackage[colorlinks=true, pdfstartview=FitV, linkcolor=purple, citecolor= purple, urlcolor=blue]{hyperref}
\renewcommand{\baselinestretch}{1.50}

\begin{document}
\title{\Large Study of $\chi _{c1,2}$ decays to $PPP$ via light meson resonances}
\author{Bo Lan$^1$, Jie Zhu$^2$, Jin-Huan Sheng$^3$, Yuan-Guo Xu$^1$, and Ru-Min Wang$^{1,\dag}$\\
{\scriptsize $^1$School of Physics, Jiangxi Normal University, Nanchang, Jiangxi 330022, China}\\
{\scriptsize $^2$School of Physics and Electrical Engineering, Anyang Normal University, Anyang,  Henan, 455000, China}\\
{\scriptsize $^3$School of Physics and Engineering, Henan University of Science and Technology, Luoyang, Henan, 471000, China}\\
	$^\dag${\scriptsize Corresponding author.~~Email:ruminwang@sina.com}
}

\begin{abstract}
The three-body decays of charmonium states provide a valuable laboratory for probing non-perturbative QCD dynamics and investigating light meson resonances.
Three-body decays  $\chi _{c1,2} \to P_{1}P_{2}P_{3}$ are studied in this work ($P$ denotes the light pseudoscalar meson).
Branching ratio analysis is performed for the $\chi _{c1,2} \to P_{1}P_{2}P_{3}$ decays  incorporating
light vector ($V$) meson resonance, light tensor ($T$) meson resonance and light scalar ($S$) meson resonance contributions. Based on  the SU(3) flavor symmetry approach, we calculate the branching ratios under both the narrow width approximation and the  finite width effect, and provide the predict results in the $\chi _{c1,2}\to  (V/T)P_{3} \to P_{1}P_{2}P_{3}$ processes. Furthermore, based on the significant $a^{+}_{0}(980)$ resonance signal in $\chi _{c1} \to \pi^{+ }\pi^{- }\eta$ decay, the predict results for $\chi _{c1}\to  SP_{3} \to P_{1}P_{2}P_{3}$ are also obtained. While the resonance signals observed in current experiments are mainly concentrated in the $\chi_{c1} \to \pi^+ \pi^- \eta$ decay, our results suggest potential resonance contributions in other channels. Especially in $\chi _{c1,2}\to KK\pi$ channels, the predicted results show the significant resonance contributions from $V$ mesons and $T$ mesons, which is worth further experimental verification and exploration.

\end{abstract}
\maketitle
	
\section{Introduction}
Charmonium, a bound state of a charm quark and its antiquark, has long been regarded as an ideal laboratory for studying the strong interaction. It provides a crucial perspective for exploring the interface between the perturbative and non-perturbative regimes of Quantum Chromodynamics (QCD) \cite{Mo:2006cy}. The spectroscopy and decay properties of charmonium offer an indispensable window for testing fundamental theories and for understanding the mechanisms of color confinement and hadronization. Among the charmonium family, the P-wave excited state $\chi_ {cJ} $($J=0,1,2 $) occupy a special position due to their positive parity and relatively rich decay modes. Their dominant hadronic decays provide an important source for investigating non-perturbative QCD dynamics \cite{Voloshin:2007dx,Vairo:2003gh}.

The systematic study of three-body $\chi _{cJ}$ decays carries profound physical significance for several reasons. First, experimental measurements of these decays provide high-precision constraints and benchmarks for testing non-perturbative QCD models, such as QCD sum rules \cite{Novikov:1976tn,Reinders:1981si,Nielsen:2009uh,Wang:2012gj,Albuquerque:2018jkn}, lattice QCD \cite{Brambilla:2019esw,Wu:2025hlf}, effective field theories \cite{Berwein:2015vca}, and the chiral unitary approach \cite{Liang:2016hmr,Li:2024uwu}. Second, these processes serve as an excellent laboratory for investigating the hadronization mechanism, and the dynamical evolution from $c\bar{c}$ pair into multiple final-state hadrons is also closely related to the nature of color confinement. Additionally, the rich resonance spectra in three-body decays may contain signals of exotic hadronic states, such as glueballs or multiquark states, and can offer unique insights into the non-perturbative challenges, such as the production of baryon-antibaryon pairs. Finally, precise knowledge of the decay properties of $\chi _{cJ}$ states is also of practical importance for background estimation, as well as for new physics searches in related processes at high energy experiments.

The $\chi_{cJ}$ states are copiously produced via the radiative transition $\Psi(2S) \to \gamma\chi_{cJ}$, providing an ideal experimental platform for studying their decay properties. With BESIII  data samples of $(107.7 \pm 0.6) \times 10^6$ $\Psi(2S)$ events from 2009 \cite{Ablikim:2012pj}, $(345.4 \pm 2.6)\times 10^6$ events taken in 2012 \cite{BESIII:2017tvm}, and $(2259.3 \pm 11.1) \times 10^6$ in 2021 \cite{BESIII:2024lks}, increasingly precise measurements have been collected, stimulating growing interest in both theoretical and experimental studies of the hadronic decays of these states \cite{BES:2005ukn,Luchinsky:2005bf,CLEO:2007rbf,CLEO:2008jma,BESIII:2010ank,Chen:2013gka,Gross:2022hyw,BESIII:2025wxa}. Despite these progresses, the full hadronic decay dynamics of the $\chi_{cJ}$ system remain poorly understood, as  many of their decay modes lacking high-precision experimental measurements and rigorous theoretical interpretations. Precise measurements of the $\chi_{cJ}$ hadronic decays provide valuable information on non-perturbative QCD dynamics and potential glueball dynamics.

Charmonium hadronic decays proceed at low momentum transfer scales, where the strong coupling constant grows significantly as a consequence of QCD asymptotic freedom, and thus non-perturbative QCD effects are  pronounced in these processes.  Theoretical studies of $\chi_{cJ}$ therefore rely on non-perturbative or phenomenological QCD approaches.  Experimentally, only resonance  branching ratios of the $\chi_{c1} \to \pi^+ \pi^- \eta$ decay channel have been reported \cite{BESIII:2016tqo}. A number of valuable theoretical predictions for related decay modes have been presented. In particular, the chiral unitary approach provides a clear physical picture and predictions \cite{Liang:2016hmr,Li:2024uwu}, while the SU(3) flavor symmetry approach offers a model-independent systematic parameterization scheme applicable to two-body, three-body, and radiative decays of charmonium states \cite{Mo:2024zsa,Haber:1985cv}.

In this work, we employ the SU(3) flavor symmetry approach to investigate resonance decay processes of the $\chi _{c1,2}\to  (V/T)P_{3} \to P_{1}P_{2}P_{3}$ and $\chi _{c1}\to  SP_{3} \to P_{1}P_{2}P_{3}$ decays. We provide systematic predictions for the resonance branching ratios of these decay channels, evaluating results  under both the narrow width approximation and with finite width effects included. Noted that, since $\chi_{c0}\to PV,PT$ decays are forbidden by spin-parity conservation, $\chi _{c0}\to  RP_{3} \to P_{1}P_{2}P_{3}$ are not discussed in this work.

This paper is organized as follows:
the theoretical framework is detailed in Sec.~\ref{sec:theory}, our numerical results and discussions are given in Sec.~\ref{sec:num}, and conclusion are presented in Sec.~\ref{sec:Conclusion}.

\section{Theoretical frame}  \label{sec:theory}

Three-body and multi-body decays offer richer dynamical information and more flexible, complementary observables compared to two-body decays. However, the corresponding theoretical calculations are more complicated, so decay information is usually obtained through the quasi-two-body approximation \cite{Capstick:1993kb,Wang:2016rlo}. With the resonance contributions considered, the three-body $\chi _{cJ} \to P_{1}P_{2}P_{3}$ decays can be factorized into the sequential processes  $\chi _{cJ}\to  RP_{3}$  followed by  $R \to P_{1}P_{2}$, where the $R$ denotes  $V$, $T$ or $S$ meson. The decay amplitude takes the form
\begin{eqnarray}	
	A=\mathcal{M}(\chi _{cJ}\to  RP_{3})R^{BW}\mathcal{M}(R\to P_{1}P_{2}),
\end{eqnarray}
where the resonance propagator is commonly described  by the relativistic Breit-Wigner line shape \cite{Cheng:2020iwk,Qi:2018lxy}
\begin{eqnarray}	
	R^{BW}=\frac{1}{(t_{R}-m^{2}_{R})+im_{R}\Gamma_{R}},	
\end{eqnarray}
where $t_{R}\to m^{2}_{R}$. After normalization, the branching ratios considering finite width effects is given by \cite{Cheng:2020iwk,Qi:2018lxy}
\begin{eqnarray}	
	\mathcal{B}(\chi _{cJ}\to  RP_{3} \to P_{1}P_{2}P_{3})=\frac{1}{\pi}\int^{(m_{R}+\Gamma_{R})^{2}}_{(m_{R}-\Gamma_{R})^{2}}\frac{\mathcal{B}(\chi _{cJ}\to  RP_{3},m_{R}\to\sqrt{t_{R}})\mathcal{B}(R\to P_{1}P_{2},m_{R}\to\sqrt{t_{R}})m_{R}\Gamma_{R}}{(t_{R}-m^{2}_{R})^{2}+m^{2}_{R}\Gamma^{2}_{R}}dt_{R}.\label{eq:FWE}
\end{eqnarray}
In the narrow width approximation, $i.e.$, $\Gamma_{R} \to 0$, Eq.~(\ref{eq:FWE}) simplifies to the product of the two sequential branching ratios
\begin{eqnarray}
	\mathcal{B}(\chi _{cJ}\to  RP_{3} \to P_{1}P_{2}P_{3})=\mathcal{B}(\chi_{cJ}\to RP_{3})\times\mathcal{B}(R \to P_{1}P_{2}). \label{eq:NWA}
\end{eqnarray}
Thus, the  three-body decay branching ratio can be directly  derived from  the independent calculations of $\mathcal{B}(\chi _{cJ}\to  RP_{3})$ and $\mathcal{B}(R \to P_{1}P_{2})$.

The branching ratios of the $\chi_{cJ}\to PV,PT$ decays can be expressed as \cite{Lan:2025wkf}
\begin{equation}
	\mathcal{B}(\chi _{cJ}\to M_1M_2)=\frac{\left\vert \vec{p}\right\vert }{8\pi M_{\chi
			_{cJ}}^{2}\Gamma _{\chi _{cJ}}}\Big| \mathcal{A}(\chi _{cJ}\to M_1M_2)\Big| ^{2}, \label{Eq:br}
\end{equation}
where $M _{\chi_{cJ}}$ and $\Gamma _{\chi_{cJ}}$ are the mass and total decay width of the parent $\chi_{cJ}$\ meson, and the modulus of the center-of-mass momentum is $|\vec{p}|\equiv\sqrt{\lambda (M_{\chi _{cJ}}^{2},m_{M_1}^2,m_{M_2}^2)}/(2M_{\chi _{cJ}})$  with $\lambda(a,b,c)=a^2+b^2+c^2-2ab-2bc-2ac$.

The $\chi _{c1}\to  PS$ decays are different from the $\chi _{cJ}\to PV,PT$ decays, and they proceed via $P$-wave.
Therefore, the momentum dependence of the $\chi _{c1}\to  PS$  channels is different from  Eq. (\ref{Eq:br})  \cite{Wang:2012qa}.
And $\mathcal{B}(\chi _{c1}\to PS)$  can be written as
\begin{equation}
\mathcal{B}(\chi _{c1}\to PS)=\frac{\left\vert \vec{p}\right\vert^{3} }{8\pi M_{\chi
			_{c1}}^{2}\Gamma _{\chi _{c1}}}\Big| \mathcal{A}(\chi _{c1}\to PS) \Big| ^{2}. \label{Eq:brps}
\end{equation}
Noted that, since there is no any $S$ resonance  relevant experimental data in $\chi _{c2}\to  PPP$, we only consider the $S$ resonances in $\chi_{c1}\to PPP$ decays.

For the $R \to  P_1P_2$ decays with $R=S/V/T$, the branching ratios are  \cite{Cheng:2020ipp,Cheng:2022ysn}
\begin{eqnarray}	
	\mathcal{B}(S \to  P_1P_2)=\frac{\tau_{S}p'_{c}}{8\pi m^{2}_{S}}g^{2}_{S \to  P_1P_2},\\
	\mathcal{B}(V \to  P_1P_2)=\frac{\tau_{V}p'^{3}_{c}}{6\pi m^{2}_{V}}g^{2}_{V \to  P_1P_2},\\
	\mathcal{B}(T \to  P_1P_2)=\frac{\tau_{T}p'^{5}_{c}}{60\pi m^{2}_{T}}g^{2}_{T \to P_1P_2},
\end{eqnarray}
where $p'_{c}\equiv\sqrt{\lambda(m^{2}_{R},m^{2}_{P_{1}},m^{2}_{P_{2}})}/(2m_{R})$,  $\tau_{R}$ is the lifetime of $R$ mesons,  and $g_{R \to  P_1P_2}$ is  strong coupling constant of $R \to  P_1P_2$ decays.

The core theoretical task of this work  is the calculation of the two body decay amplitudes, for which the SU(3) flavor symmetry approach offers a relatively simple and widely used framework for hadron decay calculations in the non-perturbative QCD regime. Within this framework, the decay amplitudes for $\chi _{cJ}\to  PV/PT$ can be parameterized as
\begin{eqnarray}	
	\mathcal{A}(\chi _{cJ}\to  M_1M_2)=a^{M}_{1J}M_{1j}^{i}M_{2i}^{j}+a^{M}_{2J}M_{1i}^{i}M_{2j}^{j}+b^{M}_{J}M_{1j}^{i}W_{k}^{j}M_{2i}^{k},\label{EQ:DA}
\end{eqnarray}
where $M=P/V/T$, the indices $(i,j,k)=1,2,3$ run over the three light quark flavors $u,d,s$, and  correspond to rows and columns of the SU(3) meson matrices, $a^{M}_{1J}$ and $a^{M}_{2J}$ are parameters that represent the SU(3) symmetry terms, and $b^{M}_{J}$ is the parameter of the breaking term.  Additionally, the breaking effects primarily arise from the differing masses of the $u$, $d$ and $s$ quarks. The diagonalized mass matrix can be expressed as \cite{Gronau:1995hm,Xu:2013dta,He:2014xha}
\begin{equation}
	\left(
	\begin{array}{ccc}
		m_{u} & 0 & 0  \\
		0 & m_{d} & 0  \\
		0 & 0 & m_{s}
	\end{array}
	\right)=\frac{1}{3}(m_{u}+m_{d}+m_{s})I + \frac{1}{2}(m_{u}-m_{d})X + \frac{1}{6}(m_{u}+m_{d}-2m_{s})W,
\end{equation}
where $I$ is the identity matrix, and $X$ and $W$ represented as
\begin{equation}
	X=\left(
	\begin{array}{ccc}
		1 & 0 & 0  \\
		0 & -1 & 0  \\
		0 & 0 & 0
	\end{array}
	\right),~~~~~~
	W=\left(
	\begin{array}{ccc}
		1 & 0 & 0  \\
		0 & 1 & 0  \\
		0 & 0 & -2
	\end{array}
	\right).
\end{equation}
Considering the mass difference between the $s$ quark and the $u$, $d$ quarks, we include only the $W$ part of the diagonalized mass matrix to account for symmetry breaking.

The relevant pseudoscalar, vector, and tensor meson  octets and singlets in the SU(3) flavor basis ($u,d,s$ quarks) can be written as \cite{Haber:1985cv,Mo:2024jjd}
\begin{eqnarray}
	P =\left(
	\begin{array}{ccc}
		\frac{\pi ^{0}}{\sqrt{2}}+\frac{\eta _{8}}{\sqrt{6}}+\frac{\eta _{1}}{\sqrt{3	}} & \pi ^{+}                                                                         & K^{+}                                                   \\
		\pi ^{-}                                                                         & -\frac{\pi ^{0}}{\sqrt{2}}+\frac{\eta _{8}}{\sqrt{6}}+\frac{\eta	_{1}}{\sqrt{3}} & K^{0}                                                   \\
		K^{-}                                                                            & \bar{K}^{0}                                                                      & -\frac{2\eta _{8}}{\sqrt{6}}+\frac{\eta _{1}}{\sqrt{3}}
	\end{array}
	\right) , \\
	V =\left(
	\begin{array}{ccc}
		\frac{\rho ^{0}}{\sqrt{2}}+\frac{\omega _{8}}{\sqrt{6}}+\frac{\omega _{1}}{	\sqrt{3}} & \rho ^{+}                                                                              & K^{\ast +}                                                  \\
		\rho ^{-}                                                                             & -\frac{\rho ^{0}}{\sqrt{2}}+\frac{\omega _{8}}{\sqrt{6}}+\frac{	\omega _{1}}{\sqrt{3}} & K^{\ast 0}                                                  \\
		K^{\ast -}                                                                            & \bar{K}^{\ast 0}                                                                       & -\frac{2\omega _{8}}{\sqrt{6}}+\frac{\omega	 _{1}}{\sqrt{3}}
	\end{array}
	\right), \\
	T =\left(
	\begin{array}{ccc}
		\frac{a ^{0}_{2}}{\sqrt{2}}+\frac{f ^{8}_{2}}{\sqrt{6}}+\frac{f^{1}_{2}}{\sqrt{3}} & a ^{+}_{2}                                                                          & K^{*+}_{2}                                              \\
		a ^{-}_{2}                                                                         & -\frac{a ^{0}_{2}}{\sqrt{2}}+\frac{f ^{8}_{2}}{\sqrt{6}}+\frac{f^{1}_{2}}{\sqrt{3}} & K^{*0}_{2}                                              \\
		K^{*-}_{2}                                                                         & \bar{K}^{*0}_{2}                                                                    & -\frac{2f^{8}_{2}}{\sqrt{6}}+\frac{f^{1}_{2}}{\sqrt{3}}
	\end{array}
	\right),
\end{eqnarray}	
where $\eta_{1}-\eta_{8}$, $\omega_{1}-\omega_{8}$, and $f^{1}_{2}-f^{8}_{2}$ are the unmixed SU(3) flavor-singlet and flavor-octet states for the pseudoscalar, vector, and tensor meson sectors, respectively.
Following the standard mixing scheme adopted by  the Particle Data Group (PDG) \cite{ParticleDataGroup:2024cfk}, the physical mixed states are
\begin{eqnarray}
	\eta  &=&\left( \eta _{8}\cos \theta _{P}-\allowbreak \eta _{1}\sin \theta	_{P}\right),  ~~~~~~~	\eta ^{\prime } =\left( \eta _{8}\sin \theta _{P}+\allowbreak \eta	_{1}\cos \theta _{P}\right),\\
\phi  &=&\left( \omega _{8}\cos \theta _{V}-\allowbreak \omega _{1}\sin \theta_{V}\right),  ~~~~~~	\omega =\left( \omega _{8}\sin \theta _{V}+\allowbreak \omega_{1}\cos \theta _{V}\right),\\
f^{'}_{2}  &=&\left( f ^{8}_{2}\cos \theta _{T}-\allowbreak f^{1}_{2}\sin \theta	_{T}\right),  ~~~~~f_{2} =\left( f ^{8}_{2}\sin \theta _{T}+\allowbreak f^{1}_{2}\cos \theta _{T}\right),
\end{eqnarray}
where $\theta _{P},~\theta _{V},~\theta _{T}$ are the nonet mixing angles for the $P$, $V$, $T$ mesons, respectively. For each mixing angle $\theta_{P,V,T}$, there are two values, one calculated from the mass of meson, and the other calculated from the square mass of meson. We adopt $\theta _{P}=[-20^\circ,-10^\circ]$ calculated from the square mass of meson, as well as $\theta _{V}=36.5^\circ$ and $\theta _{T}=(27\pm1)^\circ$ calculated from the mass of meson \cite{ParticleDataGroup:2024cfk}.
Detailed calculations and  analytical results for the decay amplitudes $\mathcal{A}(\chi _{cJ}\to  M_1M_2)$ derived from Eq.~(\ref{EQ:DA}) can be found in our previous work \cite{Lan:2025wkf}, and these results will be used for the calculation of Eq.~(\ref{eq:FWE}) and Eq.~(\ref{eq:NWA}).

In addition, for the two-body $V/T \to  P_{1}P_{2}$ decays, the strong coupling constants can be parameterized  within the SU(3) flavor symmetry framework as
\begin{eqnarray}	
	g^{s}_{M \to PP}&=&g^{M}_{10}M^{i}_{j}P^{i}_{k}P^{k}_{j}+g^{M}_{20}M^{i}_{j}P^{i}_{j}P^{k}_{k},\\
	g^{b}_{M \to PP}&=&g^{M}_{11}M^{i}_{j}W^{i}_{m}P^{m}_{k}P^{k}_{j}+g^{M}_{12}M^{i}_{j}P^{i}_{k}W^{k}_{m}P^{m}_{j}+g^{M}_{21}M^{i}_{j}W^{i}_{m}P^{m}_{j}P^{k}_{k},
\end{eqnarray}
where the superscript $s$ labels the SU(3) flavor symmetry-conserving contributions, and superscript $b$ labels the symmetry-breaking contributions induced by the strange quark mass;  $g^{M}_{10}$ and $g^{M}_{20}$ are parameters that represent the SU(3) symmetry terms, while $g^{M}_{11}$, $g^{M}_{12}$, and $g^{M}_{21}$ are the parameters of the breaking terms, the detailed calculation is similar to that of $\mathcal{A}(\chi _{cJ}\to  M_1M_2)$.

Unlike previous works that only considered symmetry-conserving terms \cite{Wang:2022fbk}, we include both symmetry-conserving and symmetry-breaking contributions in our following analysis. The detailed amplitudes are listed in Table \ref{Tab:AV2PP} for $V \to P_1P_2$ and in Table \ref{Tab:AT2PP} for $T \to P_1P_2$. By incorporating the new experimental results and symmetry-breaking terms, we are able to calculate more decay channels. The relevant numerical results will be presented in the next section.

\begin{table}[b]
	\caption{The SU(3) flavor  amplitude forms for the  $V \to P_1P_2$ decays. }
	\centering
	\begin{center}
		\renewcommand\arraystretch{1.2}\tabcolsep 0.4in
		\begin{tabular}{c|c}
			\hline\hline
			Mode               &                                                                                                                     Amplitude                                                                                                                     \\ \hline
			$\rho^{\pm}\to \pi^{\pm}\eta$   &                                                               $\Big( \frac{\cos\theta _{P}}{\sqrt{6}} -\frac{\sin\theta_{P}}{\sqrt{3}} \Big) ( g_{10}+g_{20}+g_{11}+g_{12}+g_{21})$                                                               \\
			$\rho^{0}\to \pi^{0}\eta$     &                                                             $\Big( \frac{\cos\theta _{P}}{2\sqrt{6}} -\frac{\sin\theta _{P}}{2\sqrt{3}}\Big)( g_{10}+g_{20}+g_{11}+g_{12}+g_{21})  $                                                             \\
			$\rho^{\pm}\to \pi^{\pm}\pi^{0}$ &                                                                                             $\frac{1}{\sqrt{2}}(g_{10}+g_{20}+g_{11}+g_{12}+g_{21})$                                                                                              \\
			$\rho^{0}\to \pi^{+}\pi^{-}$   &                                                                                                   $\frac{1}{\sqrt{2}}(  g_{10}+g_{11}+g_{12})$                                                                                                    \\ \hline
			$K^{*\pm}\to  K^{\pm}\pi^{0}$   &                                                                                                   $\frac{1}{\sqrt{2}}(  g_{10}+g_{11}+g_{12})$                                                                                                    \\
			$K^{*\pm}\to  K^{0}\pi^{\pm}$   &                                                                                                              $g_{10}+g_{11}+g_{12}$                                                                                                               \\
			$K^{*0}\to  K^{+}\pi^{-}$     &                                                                                                              $g_{10}+g_{11}+g_{12}$                                                                                                               \\
			$K^{*0}\to  K^{0}\pi^{0}$     &                                                                                                   $-\frac{1}{\sqrt{2}}(  g_{10}+g_{11}+g_{12})$                                                                                                   \\ \hline
			$\phi \to K^{+}K^{-}/K^{0}K^{0}$       & $-\Big( \frac{\cos\theta_{V}}{\sqrt{6}} +\frac{2\sin\theta _{V}}{\sqrt{3}} \Big) g_{10}+  \Big( \frac{5\cos\theta _{V}}{\sqrt{6}}+\frac{\sin\theta _{V}}{\sqrt{3}}\Big) g_{11}+\Big(-\frac{4\cos\theta _{V}}{\sqrt{6}} +\frac{\sin\theta _{V}}{\sqrt{3}}\Big) g_{12}$  \\
			$\phi \to \pi^{+}\pi^{-}$     & $\Big( \frac{2\cos\theta_{V}}{\sqrt{6}} -\frac{2\sin\theta _{V}}{\sqrt{3}} \Big)(g_{10}+  g_{11}+ g_{12})$  \\
			$\phi \to \pi^{0}\pi^{0}$     & $\Big( \frac{\cos\theta_{V}}{\sqrt{3}} -\frac{2\sin\theta _{V}}{\sqrt{6}} \Big) (g_{10}+  g_{11}+ g_{12})$  \\ \hline
			$\omega \to \pi^{+}\pi^{-}$    & $\Big( \frac{2\cos\theta_{V}}{\sqrt{3}} +\frac{2\sin\theta _{V}}{\sqrt{6}} \Big) (g_{10}+  g_{11}+ g_{12})$  \\
			$\omega \to \pi^{0}\pi^{0}$    & $\Big( \frac{2\cos\theta_{V}}{\sqrt{6}} +\frac{\sin\theta _{V}}{\sqrt{3}} \Big) (g_{10}+  g_{11}+ g_{12})$  \\ \hline
		\end{tabular}
	\end{center}\label{Tab:AV2PP}
	\end{table}
	\begin{table}[t]
	\caption{The SU(3) flavor  amplitude forms for the $T \to P_1P_2$ decays. 	}
	\centering
	\begin{center}
		\renewcommand\arraystretch{1.3}\tabcolsep 0.135in
		\begin{tabular}{c|c}
			\hline\hline
			Mode                &                                                                                                                                                                   Amplitude                                                                                                                                                                   \\ \hline
			$a_{2}\to \pi\eta$         & $\Big( \frac{2\cos\theta _{P}}{\sqrt{6}} -\frac{2\sin\theta _{P}}{\sqrt{3}} \Big) g_{10}  -  \sqrt{3}\sin\theta _{P}g_{20} +  \Big( \frac{2\cos\theta _{P}}{\sqrt{6}}-\frac{2\sin\theta _{P}}{\sqrt{3}} \Big)g_{11}   + \Big( \frac{2\cos\theta _{P}}{\sqrt{6}} -\frac{2\sin\theta _{P}}{\sqrt{3}}\Big) g_{12}-\sqrt{3}\sin\theta_{p} g_{21}$ \\
			$a_{2}\to \pi\eta'$        & $\Big( \frac{2\sin\theta _{P}}{\sqrt{6}} +\frac{2\cos\theta _{P}}{\sqrt{3}} \Big) g_{10}  +  \sqrt{3}\cos\theta _{P}g_{20} +  \Big( \frac{2\sin\theta _{P}}{\sqrt{6}}+\frac{2\cos\theta _{P}}{\sqrt{3}} \Big)g_{11}   + \Big( \frac{2\sin\theta _{P}}{\sqrt{6}} +\frac{2\cos\theta _{P}}{\sqrt{3}}\Big) g_{12}+\sqrt{3}\cos\theta_{p} g_{21}$ \\
			$a^{\pm}_{2}\to (K\bar{K})^{\pm}$ &                                                                                                                                                       $g_{10}+\frac{1}{2}g_{11}-g_{12}$                                                                                                                                                       \\
			$a^{0}_{2}\to  K^{0}\bar{K^{0}}$  &                                                                                                                                                 $-\frac{1}{\sqrt{2}}(g_{10}+g_{11}-2g_{12})$                                                                                                                                                  \\
			$a^{0}_{2}\to  K^{+}K^{-}$     &                                                                                                                                                  $\frac{1}{\sqrt{2}}(g_{10}+g_{11}-2g_{12})$                                                                                                                                                  \\ \hline
			$K^{*\pm}_{2}\to  K^{\pm}\pi^{0}$ &                                                                                                                                                  $\frac{1}{\sqrt{2}}(g_{10}+g_{11}+g_{12})$                                                                                                                                                   \\
			$K^{*\pm}_{2}\to  K^{0}\pi^{\pm}$ &                                                                                                                                                            $g_{10}+g_{11}+g_{12}$                                                                                                                                                             \\
			$K^{*0}_{2}\to  K^{+}\pi^{-}$   &                                                                                                                                                            $g_{10}+g_{11}+g_{12}$                                                                                                                                                             \\
			$K^{*0}_{2}\to  K^{0}\pi^{0}$   &                                                                                                                                                  $-\frac{1}{\sqrt{2}}(g_{10}+g_{11}+g_{12})$                                                                                                                                                  \\ \hline
			$f_{2} \to K\bar{K}$        &                                     $\Big( \frac{2\cos\theta_{T}}{\sqrt{3}} -\frac{\sin\theta _{T}}{\sqrt{6}} \Big) g_{10}+  \Big( -\frac{\cos\theta _{T}}{\sqrt{3}}+\frac{5\sin\theta _{T}}{\sqrt{6}}\Big) g_{11}-\Big(\frac{\cos\theta _{T}}{\sqrt{3}} +\frac{4\sin\theta _{T}}{\sqrt{6}}\Big) g_{12}$                                      \\
			$f_{2} \to \pi^+\pi^-$       &                                                                                                                  $\Big( \frac{2\cos\theta_{T}}{\sqrt{3}} +\frac{2\sin\theta _{T}}{\sqrt{6}} \Big) (g_{10}+  g_{11}+ g_{12})$                                                                                                                  \\
			$f_{2} \to \pi^0\pi^0$       &                                                                                                                  $\Big( \frac{2\cos\theta_{T}}{\sqrt{6}} +\frac{\sin\theta _{T}}{\sqrt{3}} \Big) (g_{10}+  g_{11}+ g_{12})$                                                                                                                   \\ \hline
			$f'_{2} \to K\bar{K}$       &                                     $-\Big( \frac{\cos\theta_{T}}{\sqrt{6}} +\frac{2\sin\theta _{T}}{\sqrt{3}} \Big) g_{10}+  \Big( \frac{5\cos\theta _{T}}{\sqrt{6}}+\frac{\sin\theta _{T}}{\sqrt{3}}\Big) g_{11}+\Big(-\frac{4\cos\theta _{T}}{\sqrt{6}} +\frac{\sin\theta _{T}}{\sqrt{3}}\Big) g_{12}$                                     \\
			$f'_{2} \to \pi^+\pi^-$      &                                                                                                                  $\Big( \frac{2\cos\theta_{T}}{\sqrt{6}} -\frac{2\sin\theta _{T}}{\sqrt{3}} \Big)(g_{10}+  g_{11}+ g_{12})$                                                                                                                   \\
			$f'_{2} \to \pi^0\pi^0$      &                                                                                                                  $\Big( \frac{\cos\theta_{T}}{\sqrt{3}} -\frac{2\sin\theta _{T}}{\sqrt{6}} \Big) (g_{10}+  g_{11}+ g_{12})$                                                                                                                   \\ \hline
		\end{tabular}
	\end{center}\label{Tab:AT2PP}
\end{table}

With the decay amplitudes fully derived, the free parameters can be constrained by fitting to the existing experimental data, which   allows  us to predict the  branching ratios for all relevant decay channels governed by intermediate $V$, $T$, and $S$ resonances. The detailed numerical results are presented in the following section.

\section{ Numerical results } \label{sec:num}
The theoretical input parameters, such as the lifetimes, the masses, and the experimental data within the $1\sigma$ error bar from the PDG \cite{ParticleDataGroup:2024cfk} will be used in our numerical analysis.
The branching ratios for the resonance three-body processes $\chi_{cJ} \to P_{1}P_{2}P_{3}$ can be obtained in terms of  $\mathcal{B}(\chi _{cJ}\to  PV/PT)$ and $\mathcal{B}(V/T \to  P_{1}P_{2})$. The branching ratios for the $\chi _{cJ}\to  PV/PT$ decays are taken directly from our previous SU(3) flavor analysis  \cite{Lan:2025wkf}.

\subsection{ Numerical results of $R \to  P_{1}P_{2}$ decays }

 All relevant experimental data of $V/T \to  P_{1}P_{2}$ are listed in the second columns of Tables  \ref{Tab:V2PP}-\ref{Tab:T2PP} from PDG \cite{ParticleDataGroup:2024cfk}.
Using the parametrization forms of Tables  \ref{Tab:AV2PP}-\ref{Tab:AT2PP}  and  the corresponding data of Tables  \ref{Tab:V2PP}-\ref{Tab:T2PP}, the parameters $g_{ij}$ are obtained for the different decay processes by Monte Carlo sampling method, and they are listed in Table  \ref{Tab:gij}.

\begin{table}[hptb]
	\caption{The branching ratios for the $V \to P_1P_2$ decays.}
	\centering{\footnotesize
		\begin{center}  \renewcommand\arraystretch{1.0}\tabcolsep0.3in
			\begin{tabular}{c|c|c}
				\hline\hline
				& Exp. data \cite{ParticleDataGroup:2024cfk} &                                       Predicted branching ratios                                \\ \hline
				$\mathcal{B}(\rho^{\pm}\to \pi^{\pm}\eta)$   &             $<6\times10^{-3}$              &                                       $<6\times10^{-3}$                                        \\ \hline
				$\mathcal{B}(\rho^{0}\to \pi^{0}\eta)$     &                  $\cdots$                  &                                      $<6\times10^{-3}$                                       \\ \hline
				$\mathcal{B}(\rho^{\pm}\to \pi^{\pm}\pi^{0})$ &                $\sim 100\%$                &                                          $\sim 100\%$                                           \\
				$\mathcal{B}(\rho^{0}\to \pi^{+}\pi^{-})$   &                $\sim 100\%$                &                                          $\sim 100\%$                                           \\ \hline
				$\mathcal{B}(K^{*\pm}\to  (K\pi)^{\pm})$    &      $(99.902\pm 0.009)\times10^{-2}$      &                 $(65.6\pm 1.0)\times10^{-2}$$_{[K^{*\pm}\to  K^{0}\pi^{\pm}]}$                  \\
				&                                            &                 $(34.3\pm 1.0)\times10^{-2}$$_{[K^{*\pm}\to  K^{\pm}\pi^{0}]}$                  \\ \hline
				$\mathcal{B}(K^{*0}\to  (K\pi)^{0})$      &      $(99.754\pm 0.021)\times10^{-2}$      &    $(66.7\pm 1.0)\times10^{-2}$$_{[K^{*0}\to  K^{+}\pi^{-},\bar{K}^{*0}\to  K^{-}\pi^{+}]}$     \\
				&                                            & $(33.0\pm 1.0)\times10^{-2}$$_{[K^{*0}\to  K^{0}\pi^{0},\bar{K}^{*0}\to  \bar{K^{*0}}\pi^{0}]}$ \\ \hline
				$\mathcal{B}(\phi \to K^{+}K^{-})$       &        $(49.9\pm 0.2)\times10^{-2}$        &                                  $(49.9\pm 0.2)\times10^{-2}$                                   \\ \hline
				$\mathcal{B}(\phi \to K^{0}K^{0})$       &                  $\cdots$                  &                                  $(32.9\pm 0.3)\times10^{-2}$                                   \\ \hline
				$\mathcal{B}(\phi \to \pi^{+}\pi^{-})$     &        $(9.5\pm 1.9)\times10^{-5}$         &                                   $(9.5\pm 1.9)\times10^{-5}$                                   \\ \hline
				$\mathcal{B}(\phi \to \pi^{0}\pi^{0})$     &                  $\cdots$                  &                                  $(4.79\pm 0.95)\times10^{-5}$                                  \\ \hline
				$\mathcal{B}(\omega \to \pi^{+}\pi^{-})$    &       $(1.53\pm 0.12)\times10^{-2}$        &                                  $(1.53\pm 0.12)\times10^{-2}$                                  \\ \hline
				$\mathcal{B}(\omega \to \pi^{0}\pi^{0})$    &            $<2.2\times10^{-4}$             &                                       $<2.2\times10^{-4}$                                       \\ \hline
			\end{tabular}
	\end{center}}\label{Tab:V2PP}\vspace{1cm}
	\caption{The branching ratios for the $T \to P_1P_2$ decays. 	}
	\centering{\footnotesize
		\begin{center}  \renewcommand\arraystretch{1.0}\tabcolsep0.3in
			\begin{tabular}{c|c|c}
				\hline\hline
				& Exp. data  \cite{ParticleDataGroup:2024cfk} &                                                  Predicted branching ratios                                                 \\ \hline
				$\mathcal{B}(a_{2}\to \pi\eta)$         &        $(14.5\pm 1.2)\times10^{-2}$         &                                      $(14.5\pm 1.2)\times10^{-2}$                                       \\ \hline
				$\mathcal{B}(a_{2}\to \pi\eta')$        &         $(5.5\pm 0.9)\times10^{-3}$         &                                       $(5.5\pm 0.9)\times10^{-3}$                                       \\ \hline
				$\mathcal{B}(a^{\pm}_{2}\to (K\bar{K})^{\pm})$ &         $(4.9\pm 0.8)\times10^{-2}$         &                                       $(4.9\pm 0.8)\times10^{-2}$                                       \\ \hline
				$\mathcal{B}(a^{0}_{2}\to (K\bar{K})^{0})$   &         $(4.9\pm 0.8)\times10^{-2}$         &                   $(2.39\pm 0.39)\times10^{-2}$$_{[a^{0}_{2}\to  K^{0}\bar{K^{0}}]}$                    \\
				&                                             &                      $(2.51\pm 0.41)\times10^{-2}$$_{[a^{0}_{2}\to  K^{+}K^{-}]}$                       \\ \hline
				$\mathcal{B}(K^{*\pm}_{2}\to  (K\pi)^{\pm})$  &        $(49.9\pm 1.2)\times10^{-3}$         &                   $(33.1\pm 0.8)\times10^{-3}$$_{[K^{*\pm}_{2}\to  K^{0}\pi^{\pm}]}$                    \\
				&                                             &                   $(16.8\pm 0.4)\times10^{-3}$$_{[K^{*\pm}_{2}\to  K^{\pm}\pi^{0}]}$                    \\ \hline
				$\mathcal{B}(K^{*0}_{2}\to  (K\pi)^{0})$    &        $(49.9\pm 1.2)\times10^{-3}$         &    $(33.3\pm 0.8)\times10^{-3}$$_{[K^{*0}_{2}\to  K^{+}\pi^{-},\bar{K^{*0}}_{2}\to  K^{-}\pi^{+}]}$     \\
				&                                             & $(16.6\pm 0.4)\times10^{-3}$$_{[K^{*0}_{2}\to  K^{0}\pi^{0},\bar{K^{*0}}_{2}\to  \bar{K^{*0}}\pi^{0}]}$ \\ \hline
				$\mathcal{B}(f_{2} \to K\bar{K})$        &         $(4.6\pm 0.4)\times10^{-2}$         &                        $(2.37\pm 0.21)\times10^{-2}$$_{[f_{2}\to  K^{+}K^{-}]}$                         \\
				&                                             &                     $(2.23\pm 0.19)\times10^{-2}$$_{[f_{2}\to  K^{0}\bar{K^{0}}]}$                      \\ \hline
				$\mathcal{B}(f_{2} \to \pi\pi)$         &      $84.3^{+2.8}_{-1.0}\times10^{-2}$      &                       $(56.6\pm 1.3)\times10^{-2}$$_{[f_{2}\to  \pi^{+}\pi^{-}]}$                       \\
				&                                             &                       $(28.6\pm 0.6)\times10^{-2}$$_{[f_{2}\to \pi^{0}\pi^{0}]}$                        \\ \hline
				$\mathcal{B}(f_{2} \to \eta\eta)$        &         $(4.0\pm 0.8)\times10^{-3}$         &                                       $(4.0\pm 0.8)\times10^{-3}$                                       \\ \hline
				$\mathcal{B}(f'_{2} \to K\bar{K})$       &        $(88.8\pm 2.2)\times10^{-2}$         &                        $(45.1\pm 1.1)\times10^{-2}$$_{[f'_{2}\to  K^{+}K^{-}]}$                         \\
				&                                             &                     $(43.7\pm 1.1)\times10^{-2}$$_{[f'_{2}\to  K^{0}\bar{K^{0}}]}$                      \\ \hline
				$\mathcal{B}(f'_{2} \to \pi\pi)$        &         $(8.2\pm 1.5)\times10^{-3}$         &                       $(5.5\pm 1.0)\times10^{-3}$$_{[f'_{2}\to  \pi^{+}\pi^{-}]}$                       \\
				&                                             &                       $(2.7\pm 0.5)\times10^{-3}$$_{[f'_{2}\to \pi^{0}\pi^{0}]}$                        \\ \hline
				$\mathcal{B}(f'_{2} \to \eta\eta)$       &        $(10.3\pm 2.2)\times10^{-2}$         &                                      $(10.3\pm 2.2)\times10^{-2}$                                       \\ \hline
			\end{tabular}
	\end{center}}\label{Tab:T2PP}
\end{table}

\begin{table}[hptb]
	\caption{The SU(3) parameters for the $V/T \to P_1P_2$ decays. 	}
	\centering{\footnotesize
		\begin{center}  \renewcommand\arraystretch{1.5}\tabcolsep0.3in
			\begin{tabular}{c|ccc|cc}
				\hline\hline
				&    $g_{10}$    &    $g_{11}$    &    $g_{12}$     &    $g_{20}$    &    $g_{21}$    \\ \hline
				$\rho$    & $12.18\pm6.60$ & $-0.82\pm8.81$ &  $0.09\pm7.44$  & $3.31\pm3.30$ & $-0.49\pm4.42$ \\
				$K^{*}$   & $4.40\pm1.08$  & $-0.15\pm2.04$ &  $0.04\pm2.89$  &                &                \\
				$\phi$    & $3.10\pm0.52$  & $0.01\pm1.57$  & $-0.03\pm1.94$  &                &                \\
				$\omega$   & $2.50\pm2.44$  & $0.02\pm2.61$  & $-0.04\pm2.25$  &                &                \\ \hline
				$a_{2}$   & $11.32\pm3.70$ & $-0.16\pm7.60$ &  $0.01\pm4.67$  & $3.06\pm3.06$  & $-0.03\pm3.01$ \\
				$K^{*}_{2}$ & $4.01\pm1.53$  & $0.02\pm0.62$  & $-0.29\pm1.61$  &                &                \\
				$f$     & $16.07\pm6.86$ & $0.12\pm11.61$ &  $0.01\pm7.12$  &                &                \\
				$f'$     & $13.66\pm7.93$ & $-0.05\pm6.04$ & $-0.54\pm11.05$ &                &                \\ \hline
			\end{tabular}
	\end{center}}\label{Tab:gij}
\end{table}

Then using these bounded parameters to calculate the corresponding branching ratios of the $V/T \to  P_{1}P_{2}$ decays.
 For each decay process, a large number of random samples are generated to ensure that the distribution of numerical results contains at least 10,000 valid sets. The numerical results  of $\mathcal{B}(V \to  P_{1}P_{2})$ and $\mathcal{B}(T \to  P_{1}P_{2})$ are presented in the last columns of Table \ref{Tab:V2PP} and Table \ref{Tab:T2PP}, respectively.

For comparison with previous coupling results \cite{Cheng:2020ipp,Wang:2022fbk}, the strong coupling constants for $V \to P_1P_2$ and $T \to P_1P_2$ are determined from  the obtained branching ratios in the last  columns of Table \ref{Tab:V2PP} and Table \ref{Tab:T2PP}. Some of them are shown as follows.
\begin{equation}	
\begin{aligned}
		&|g_{\rho^{\pm} \to \pi^{\pm}\pi^{0}}|=5.94\pm0.03, &&|g_{\rho^{\pm} \to \pi^{\pm}\eta}|=1.55\pm0.09,\\
		&|g_{K^{*\pm}\to  K^{\pm}\pi^{0}}|=3.37\pm0.08, &&|g_{K^{*0}\to  K^{0}\pi^{0}}|=3.12\pm0.06,\\
		&|g_{\phi\to  K^{+} K^{-}}|=4.51\pm0.03, &&|g_{\phi\to  \pi^{0} \pi^{0}}|=0.0058\pm0.0006,\\
		&|g_{\omega\to  \pi^{+} \pi^{-}}|=0.177\pm0.008, &&|g_{\omega\to  \pi^{0} \pi^{0}}|=0.011\pm0.0010,
\end{aligned}
\end{equation}
and
\begin{equation}	
	\begin{aligned}
		&|g_{a_{2} \to \pi\eta}|=10.74\pm0.69, &&|g_{a_{2} \to \pi\eta'}|=9.84\pm1.05,\\
		&|g_{a^{\pm}_{2}\to (K\bar{K})^{\pm}}|=10.37\pm1.09,&&|g_{a^{0}_{2}\to  K^{0}\bar{K}^{0}}|=7.42\pm0.78, \\
		&|g_{K^{*\pm}_{2}\to  K^{\pm}\pi^{0}}|=2.70\pm0.12,&&|g_{K^{*0}_{2}\to  K^{0}\pi^{0}}|=2.74\pm0.10, \\
		&|g_{f_{2}\to  K^{+} K^{-}}|=11.25\pm0.59,  &&|g_{f_{2}\to  \pi^{0} \pi^{0}}|=13.20\pm0.24,\\
		&|g_{f_{2}\to  \eta \eta}|=7.85\pm0.87, &&|g_{f'_{2}\to  \eta \eta}|=9.01\pm1.39,\\
		&|g_{f'_{2}\to  K^{+} K^{-}}|=14.94\pm0.90, &&|g_{f'_{2}\to  \pi^{0} \pi^{0}}|=0.61\pm0.08.
	\end{aligned}
\end{equation}
Other strong coupling constants can be obtain by the following relationships.
\begin{equation}	
	\begin{aligned}
	 &|g_{\rho^{0} \to \pi^{+}\pi^{-}}|=|g_{\rho^{\pm} \to \pi^{\pm}\pi^{0}}|,   &&|g_{\rho^{0} \to \pi^{0}\eta}|=|g_{\rho^{\pm} \to \pi^{\pm}\eta}|,\\
	 &|g_{K^{*\pm}\to  K^{0}\pi^{\pm}}|=\sqrt{2}|g_{K^{*\pm}\to  K^{\pm}\pi^{0}}|, &&|g_{K^{*0}\to  K^{+}\pi^{-}}|=\sqrt{2}|g_{K^{*0}\to  K^{0}\pi^{0}}|,	\\
	 &|g_{\phi\to  K^{0} K^{0}}|=|g_{\phi\to  K^{+} K^{-}}|,	&&|g_{\phi\to  \pi^{+} \pi^{-}}|=\sqrt{2}|g_{\phi\to  \pi^{0} \pi^{0}}|,		
	\end{aligned}
\end{equation}
and
\begin{equation}	
	\begin{aligned}
	&|g_{a^{0}_{2}\to (K\bar{K})^{0}}|=|g_{a^{\pm}_{2}\to (K\bar{K})^{\pm}}|,	 &&|g_{a^{0}_{2}\to  K^{+}K^{-}}|=|g_{a^{0}_{2}\to  K^{0}\bar{K}^{0}}|,\\
	 &|g_{K^{*\pm}_{2}\to  K^{0}\pi^{\pm}}|=\sqrt{2}|g_{K^{*\pm}_{2}\to  K^{\pm}\pi^{0}}|,	&&|g_{K^{*0}_{2}\to  K^{+}\pi^{-}}|=\sqrt{2}|g_{K^{*0}_{2}\to  K^{0}\pi^{0}}|,\\
		&|g_{f_{2}\to  K^{0} K^{0}}|=|g_{f_{2}\to  K^{+} K^{-}}|,	&&|g_{f_{2}\to  \pi^{+} \pi^{-}}|=\sqrt{2}|g_{f_{2}\to  \pi^{0} \pi^{0}}|, \\
	 &|g_{f'_{2}\to  K^{0} K^{0}}|=|g_{f'_{2}\to  K^{+} K^{-}}|,  &&|g_{f'_{2}\to  \pi^{+} \pi^{-}}|=\sqrt{2}|g_{f'_{2}\to  \pi^{0} \pi^{0}}|.
	\end{aligned}
\end{equation}

As for $S \to P_1P_2$ decays, since  the situation of $S$ mesons is more complex, and there is very little experimental data. The results  of the $S \to P_1P_2$ decays still refers to the previous results ($S$ as a two-quark state) \cite{Wang:2022fbk}.

\subsection{ Branching ratios of $\chi _{c1,2}\to RP_{3}\to P_{1}P_{2}P_{3}$ }

We calculate the three-body decay branching ratios of $\chi_{cJ}$ mesons under two scenarios: the branching ratios under the narrow width approximation, obtained from Eq.~(\ref{eq:NWA}), are shown in the Tables \ref{Tab:Brchi12PPP}-\ref{Tab:Brchi22PPP};  with finite width effects  taken into account via Eq.~(\ref{eq:FWE}), the recalculated branching ratios are also presented in the Tables \ref{Tab:Brchi12PPP}-\ref{Tab:Brchi22PPP}. For comparison, the experimental data for $\chi _{c1,2}\to   P_{1}P_{2}P_{3}$ decays are listed in second columns of Tables \ref{Tab:Brchi12PPP}-\ref{Tab:Brchi22PPP}.

\begin{sidewaystable}[htbp]
	\caption{Branching ratios of the $\chi _{c1} \to P_{1}P_{2}P_{3}$ decays with the scalar, vector, and tensor resonances. NWA represents the results under the narrow width approximation, and FWE represents the results under the finite width effect. }
	\centering
	\renewcommand\arraystretch{1.4}\tabcolsep 0.02in
	\resizebox{\textwidth}{!}{
		\begin{tabular}{c|l|l l|l l|l l}
			\hline\hline
			                 Decay modes                  & Exp.data   \cite{ParticleDataGroup:2024cfk}                                             & $S$ Resonance (NWA)                                              & $S$ Resonance (FWE)                                              & $V$ Resonance (NWA)                                            & $V$ Resonance (FWE)                                            & $T$ Resonance (NWA)                                                 & $T$ Resonance (FWE)                                                 \\ \hline
			                                              & $(4.62\pm0.24)\times10^{-3}$                                                            & $(3.20\pm0.40)\times10^{-3}_{[a_0(980)^{\pm}\to \pi^{\pm}\eta]}$ & $(3.20\pm0.40)\times10^{-3}_{[a_0(980)^{\pm}\to \pi^{\pm}\eta]}$ & $<1.62\times10^{-5}_{[\rho^{\pm}\to \pi^{\pm}\eta]}$           & $<1.42\times10^{-5}_{[\rho^{\pm}\to \pi^{\pm}\eta]}$           & $(2.31\pm1.72)\times10^{-4}_{[a^{\pm}_{2} \to \pi^{\pm}\eta]}$      & $(1.72\pm1.31)\times10^{-4}_{[a^{\pm}_{2} \to \pi^{\pm}\eta]}$      \\
			    $\chi _{c1} \to \pi^{+ }\pi^{- }\eta$     & $(1.76\pm0.24)\times10^{-4}_{[a^{\pm}_{2} \to \pi^{\pm}\eta]}$                          & $<1.51\times10^{-4}_{[f_0(980)\to \pi^{+}\pi^-]}$                & $<1.23\times10^{-4}_{[f_0(980)\to \pi^{+}\pi^-]}$                & $<9.08\times10^{-8}_{[\phi\to \pi^{+}\pi^{-}]}$                & $<6.40\times10^{-8}_{[\phi\to \pi^{+}\pi^{-}]}$                & $(3.81\pm0.70)\times10^{-4}_{[f_{2}\to \pi^{+}\pi^{-}]}$            & $(2.72\pm0.50)\times10^{-4}_{[f_{2}\to \pi^{+}\pi^{-}]}$            \\
			                                              & $(3.5\pm0.6)\times10^{-4}_{[f_{2}\to \pi^{+}\pi^{-}]}$                                  & $(3.38\pm0.75)\times10^{-3}_{[f_0(500)\to \pi^{+}\pi^-]}$        & $(2.00\pm0.46)\times10^{-3}_{[f_0(500)\to \pi^{+}\pi^-]}$        & $<3.81\times10^{-5}_{[\omega\to \pi^{+}\pi^{-}]}$      & $<2.69\times10^{-5}_{[\omega\to \pi^{+}\pi^{-}]}$      & $(2.58\pm2.09)\times10^{-6}_{[f'_{2}\to \pi^{+}\pi^{-}]}$           & $(1.83\pm1.48)\times10^{-6}_{[f'_{2}\to \pi^{+}\pi^{-}]}$           \\
			                                              & $(3.2\pm0.4)\times10^{-3}_{[a_{0}(980)^{\pm} \to \pi^{\pm}\eta]}$ \cite{BESIII:2016tqo} &                                                                  &                                                                  &                                                                &                                                                &                                                                     &                                                                     \\ \hline
			                                              &                                                                                         & $(1.73\pm0.53)\times10^{-3}_{[f_0(980)\to \pi^{+}\pi^-]}$        & $(1.51\pm0.50)\times10^{-3}_{[f_0(980)\to \pi^{+}\pi^-]}$        & $<1.96\times10^{-7}_{[\phi\to \pi^{+}\pi^{-}]}$                & $<1.39\times10^{-7}_{[\phi\to \pi^{+}\pi^{-}]}$                & $(9.27\pm7.18)\times10^{-6}_{[a^{\pm}_{2} \to \pi^{\pm}\eta']}$     & $(7.47\pm5.99)\times10^{-6}_{[a^{\pm}_{2} \to \pi^{\pm}\eta']}$     \\
			   $\chi _{c1} \to \pi^{+ }\pi^{- }\eta'$     & $(2.2\pm0.4)\times10^{-3}$                                                              & $<1.88\times10^{-4}_{[f_0(500)\to \pi^{+}\pi^-]}$                & $<1.14\times10^{-4}_{[f_0(500)\to \pi^{+}\pi^-]}$                & $<6.34\times10^{-5}_{[\omega\to \pi^{+}\pi^{-}]}$              & $<4.48\times10^{-5}_{[\omega\to \pi^{+}\pi^{-}]}$              & $<8.84\times10^{-4}_{[f_{2} \to \pi^{+}\pi^{-}]}$                   & $<6.30\times10^{-4}_{[f_{2} \to \pi^{+}\pi^{-}]}$                   \\
			                                              &                                                                                         &                                                                  &                                                                  &                                                                &                                                                & $<9.76\times10^{-6}_{[f'_{2} \to \pi^{+}\pi^{-}]}$                  & $<6.88\times10^{-6}_{[f'_{2} \to \pi^{+}\pi^{-}]}$                  \\ \hline
			  $\chi _{c1} \to \pi^{+ }\pi^{- }\pi^{0}$    & $\cdots$                                                                                & $\cdots$                                                         & $\cdots$                                                         & $(1.39\pm1.32)\times10^{-3}_{[\rho^{\pm}\to\pi^{\pm}\pi^{0}]}$ & $(9.78\pm9.32)\times10^{-4}_{[\rho^{\pm}\to\pi^{\pm}\pi^{0}]}$ & $\cdots$                                                            & $\cdots$                                                            \\
			                                              &                                                                                         &                                                                  &                                                                  & $(1.39\pm1.32)\times10^{-3}_{[\rho^{0}\to\pi^{+}\pi^{-}]}$     & $(9.77\pm9.27)\times10^{-4}_{[\rho^{0}\to\pi^{+}\pi^{-}]}$     &                                                                     &                                                                     \\ \hline
			                                              &                                                                                         & $(4.07\pm0.61)\times10^{-3}_{[a_0(980)^0\to \pi^0\eta]}$         & $(3.20\pm0.40)\times10^{-3}_{[a_0(980)^0\to \pi^0\eta]}$         & $<1.61\times10^{-5}_{[\rho^{0}\to \pi^{0}\eta]}$               & $<1.37\times10^{-5}_{[\rho^{0}\to \pi^{0}\eta]}$               & $(2.32\pm1.72)\times10^{-4}_{[a^{0}_{2}\to \pi^{0}\eta]}$           & $(1.72\pm1.33)\times10^{-4}_{[a^{0}_{2}\to \pi^{0}\eta]}$           \\
			     $\chi _{c1} \to \pi^{0}\pi^{0}\eta$      & $\cdots$                                                                                & $<7.56\times10^{-5}_{[f_0(980)\to \pi^0\pi^0]}$                & $<6.17\times10^{-5}_{[f_0(980)\to \pi^0\pi^0]}$                & $<4.58\times10^{-8}_{[\phi\to \pi^{0}\pi^{0}]}$                & $<3.22\times10^{-8}_{[\phi\to \pi^{0}\pi^{0}]}$                & $(1.92\pm0.35)\times10^{-4}_{[f_{2}\to \pi^{0}\pi^{0}]}$            & $(1.37\pm0.25)\times10^{-4}_{[f_{2}\to \pi^{0}\pi^{0}]}$            \\
			                                              &                                                                                         & $(1.73\pm0.40)\times10^{-3}_{[f_0(500)\to \pi^0\pi^0]}$          & $(1.02\pm0.25)\times10^{-3}_{[f_0(500)\to \pi^0\pi^0]}$          & $<5.02\times10^{-7}_{[\omega\to \pi^{0}\pi^{0}]}$              & $<3.56\times10^{-7}_{[\omega\to \pi^{0}\pi^{0}]}$              & $(1.31\pm1.05)\times10^{-6}_{[f'_{2}\to \pi^{0}\pi^{0}]}$           & $(0.92\pm0.75)\times10^{-6}_{[f'_{2}\to \pi^{0}\pi^{0}]}$           \\ \hline
			                                              &                                                                                         & $(8.68\pm2.68)\times10^{-4}_{[f_0(980)\to \pi^0\pi^0]}$          & $(7.57\pm2.51)\times10^{-4}_{[f_0(980)\to \pi^0\pi^0]}$          & $<9.92\times10^{-8}_{[\phi\to \pi^{0}\pi^{0}]}$                & $<7.02\times10^{-8}_{[\phi\to \pi^{0}\pi^{0}]}$                & $(9.26\pm7.20)\times10^{-6}_{[a^{0}_{2}\to \pi^{0}\eta']}$          & $(7.75\pm6.19)\times10^{-6}_{[a^{0}_{2}\to \pi^{0}\eta']}$          \\
			    $\chi _{c1} \to \pi^{0}\pi^{0}\eta'$      & $\cdots$                                                                                & $<9.60\times10^{-5}_{[f_0(500)\to \pi^0\pi^0]}$                  & $<5.77\times10^{-5}_{[f_0(500)\to \pi^0\pi^0]}$                  & $<8.44\times10^{-7}_{[\omega\to \pi^{0}\pi^{0}]}$              & $<6.32\times10^{-7}_{[\omega\to \pi^{0}\pi^{0}]}$              & $<4.44\times10^{-4}_{[f_{2}\to \pi^{0}\pi^{0}]}$                    & $<3.18\times10^{-4}_{[f_{2}\to \pi^{0}\pi^{0}]}$                    \\
			                                              &                                                                                         &                                                                  & $<4.03\times10^{-7}_{[a_0(980)^0\to \pi^0\pi^0]}$             &                                                                &                                                                & $<4.92\times10^{-6}_{[f'_{2}\to \pi^{0}\pi^{0}]}$                   & $<3.44\times10^{-6}_{[f'_{2}\to \pi^{0}\pi^{0}]}$                   \\ \hline
			     $\chi _{c1} \to K^{+}K^{-}\pi^{0}$       & $(1.81\pm0.24)\times10^{-3}$                                                            & $(4.31\pm1.23)\times10^{-4}_{[a_0(980)^0\to K^+K^-]}$            & $(2.03\pm1.30)\times10^{-4}_{[a_0(980)^0\to K^+K^-]}$            & $(1.99\pm0.41)\times10^{-4}_{[K^{*\pm}\to K^{\pm}\pi^{0}]}$    & $(1.41\pm0.29)\times10^{-4}_{[K^{*\pm}\to K^{\pm}\pi^{0}]}$    & $(4.23\pm3.29)\times10^{-5}_{[a^{0}_{2}\to K^{+}K^{-}]}$            & $(3.21\pm2.55)\times10^{-5}_{[a^{0}_{2}\to K^{+}K^{-}]}$            \\
			                                              &                                                                                         & $(1.61\pm0.28)\times10^{-3}_{[K_0^\pm\to K^\pm\pi^0]}$           & $(9.47\pm1.73)\times10^{-4}_{[K_0^\pm\to K^\pm\pi^0]}$           &                                                                &                                                                & $(1.42\pm0.35)\times10^{-5}_{[K^{*\pm}_{2}\to K^{\pm}\pi^{0}]}$     & $(0.99\pm0.23)\times10^{-5}_{[K^{*\pm}_{2}\to K^{\pm}\pi^{0}]}$     \\ \hline
			                                              &                                                                                         & $(1.57\pm0.41)\times10^{-3}_{[a_0(980)^\pm\to K^0K^\pm]}$        & $(3.66\pm2.38)\times10^{-4}_{[a_0(980)^\pm\to K^0K^\pm]}$        & $(8.36\pm1.80)\times10^{-4}_{[K^{*+}\to K^{0}\pi^{+}+c.c.]}$   & $(5.88\pm1.28)\times10^{-4}_{[K^{*+}\to K^{0}\pi^{+}+c.c.]}$   & $(1.65\pm1.28)\times10^{-4}_{[a^{+}_{2}\to K^{+}\bar{K}^{0}+c.c.]}$ & $(1.22\pm0.98)\times10^{-4}_{[a^{+}_{2}\to K^{+}\bar{K}^{0}+c.c.]}$ \\
			$\chi _{c1} \to K^{+}\bar{K}^{0}\pi^{-}+c.c.$ & $(7.0\pm0.6)\times10^{-3}$                                                              & $(3.18\pm0.56)\times10^{-3}_{[K_0^0\to K^+\pi^-+c.c.]}$          & $(1.86\pm0.34)\times10^{-3}_{[K_0^0\to K^+\pi^-+c.c.]}$          & $(6.88\pm1.10)\times10^{-4}_{[K^{*0}\to K^{+}\pi^{-}+c.c.]}$   & $(4.86\pm0.77)\times10^{-4}_{[K^{*0}\to K^{+}\pi^{-}+c.c.]}$   & $(5.32\pm1.32)\times10^{-5}_{[K^{*+}_{2}\to K^{0}\pi^{+}+c.c.]}$    & $(3.78\pm0.88)\times10^{-5}_{[K^{*+}_{2}\to K^{0}\pi^{+}+c.c.]}$    \\
			                                              &                                                                                         & $(3.16\pm0.56)\times10^{-3}_{[K^{*+}_{0}\to K^{0}\pi^{+}+c.c.]}$ & $(1.85\pm0.34)\times10^{-3}_{[K^{*+}_{0}\to K^{0}\pi^{+}+c.c.]}$ &                                                                &                                                                & $(3.92\pm0.75)\times10^{-5}_{[K^{*0}_{2}\to K^{+}\pi^{-}+c.c.]}$    & $(2.78\pm0.54)\times10^{-5}_{[K^{*0}_{2}\to K^{+}\pi^{-}+c.c.]}$    \\ \hline
			  $\chi _{c1} \to K^{0}\bar{K^{0}}\pi^{0}$    & $\cdots$                                                                                & $(2.65\pm0.76)\times10^{-4}_{[a_0(980)^0\to K^0\bar{K}^0]}$      & $(1.64\pm1.09)\times10^{-4}_{[a_0(980)^0\to K^0\bar{K}^0]}$      & $(1.71\pm0.30)\times10^{-4}_{[K^{*0}\to K^{0}\pi^{0}+c.c.]}$   & $(1.21\pm0.21)\times10^{-4}_{[K^{*0}\to K^{0}\pi^{0}+c.c.]}$   & $(4.01\pm3.11)\times10^{-5}_{[a^{0}_{2}\to K^{0}K^{0}]}$            & $(3.06\pm2.44)\times10^{-5}_{[a^{0}_{2}\to K^{0}K^{0}]}$            \\
			                                              &                                                                                         & $(1.59\pm0.28)\times10^{-3}_{[K_0^0\to K^0\pi^0+c.c.]}$          & $(9.29\pm1.69)\times10^{-4}_{[K_0^0\to K^0\pi^0+c.c.]}$          &                                                                &                                                                & $(9.72\pm1.87)\times10^{-6}_{[K^{*0}_{2}\to K^{0}\pi^{0}+c.c.]}$    & $(6.89\pm1.33)\times10^{-6}_{[K^{*0}_{2}\to K^{0}\pi^{0}+c.c.]}$    \\ \hline
			                                              &                                                                                         &                                                                  &                                                                  &                                                                &                                                                & $(2.50\pm2.17)\times10^{-6}_{[K^{*\pm}_{2}\to K^{\pm}\eta]}$        & $(1.81\pm1.56)\times10^{-6}_{[K^{*\pm}_{2}\to K^{\pm}\eta]}$        \\
			      $\chi _{c1} \to \eta K^{+}K^{-}$        & $(3.2\pm1.0)\times10^{-4}$                                                              & $<5.30\times10^{-5}_{[f_0(980)\to K^+K^-]}$                    & $<2.48\times10^{-5}_{[f_0(980)\to K^+K^-]}$                    & $<4.02\times10^{-4}_{[\phi\to K^{+}K^{-}]}$                    & $<2.84\times10^{-4}_{[\phi\to K^{+}K^{-}]}$                    & $(1.61\pm0.39)\times10^{-5}_{[f_{2}\to K^{+}K^{-}]}$                & $(1.28\pm0.31)\times10^{-5}_{[f_{2}\to K^{+}K^{-}]}$                \\
			                                              &                                                                                         &                                                                  &                                                                  &                                                                &                                                                & $(1.92\pm1.44)\times10^{-4}_{[f'_{2}\to K^{+}K^{-}]}$               & $(1.36\pm1.02)\times10^{-4}_{[f'_{2}\to K^{+}K^{-}]}$               \\ \hline
			                                              &                                                                                         &                                                                  &                                                                  &                                                                &                                                                & $(1.80\pm1.55)\times10^{-6}_{[K^{*0}_{2}\to K^{0}\eta+c.c.]}$       & $(1.30\pm1.12)\times10^{-6}_{[K^{*0}_{2}\to K^{0}\eta+c.c.]}$       \\
			   $\chi _{c1} \to \eta K^{0}\bar{K}^{0}$     & $\cdots$                                                                                & $<4.28\times10^{-5}_{[f_0(980)\to K^0\bar{K}^0]}$                & $<1.92\times10^{-5}_{[f_0(980)\to K^0\bar{K}^0]}$              & $<2.64\times10^{-4}_{[\phi\to K^{0}K^{0}]}$                    & $<1.86\times10^{-4}_{[\phi\to K^{0}K^{0}]}$                    & $(1.51\pm0.37)\times10^{-5}_{[f_{2}\to K^{0}\bar{K}^{0}]}$          & $(1.22\pm0.30)\times10^{-5}_{[f_{2}\to K^{0}\bar{K}^{0}]}$          \\
			                                              &                                                                                         &                                                                  &                                                                  &                                                                &                                                                & $(1.87\pm1.40)\times10^{-4}_{[f'_{2}\to K^{0}\bar{K}^{0}]}$         & $(1.32\pm0.99)\times10^{-4}_{[f'_{2}\to K^{0}\bar{K}^{0}]}$         \\ \hline
			      $\chi _{c1} \to \eta' K^{+}K^{-}$       & $(8.8\pm0.9)\times10^{-4}$                                                              & $(5.40\pm2.85)\times10^{-4}_{[f_0(980)\to K^{+}K^-]}$            & $(2.60\pm2.20)\times10^{-4}_{[f_0(980)\to K^{+}K^-]}$            & $<8.76\times10^{-4}_{[\phi\to K^{+}K^{-}]}$                    & $<6.18\times10^{-4}_{[\phi\to K^{+}K^{-}]}$                    & $<3.92\times10^{-5}_{[f_{2}\to K^{+}K^{-}]}$                        & $<3.12\times10^{-5}_{[f_{2}\to K^{+}K^{-}]}$                        \\
			                                              &                                                                                         &                                                                  &                                                                  &                                                                &                                                                & $<7.00\times10^{-4}_{[f'_{2}\to K^{+}K^{-}]}$                       & $<4.96\times10^{-4}_{[f'_{2}\to K^{+}K^{-}]}$                       \\ \hline
			   $\chi _{c1} \to \eta' K^{0}\bar{K^{0}}$    & $\cdots$                                                                                & $(4.36\pm2.30)\times10^{-4}_{[f_0(980)\to K^0\bar{K}^0]}$        & $(1.99\pm1.75)\times10^{-4}_{[f_0(980)\to K^0\bar{K}^0]}$        & $<5.72\times10^{-4}_{[\phi\to K^{0}K^{0}]}$                    & $<4.04\times10^{-4}_{[\phi\to K^{0}K^{0}]}$                    & $<3.70\times10^{-5}_{[f_{2}\to K^{0}\bar{K}^{0}]}$                  & $<2.96\times10^{-5}_{[f_{2}\to K^{0}\bar{K}^{0}]}$                  \\
			                                              &                                                                                         &                                                                  &                                                                  &                                                                &                                                                & $<6.80\times10^{-4}_{[f'_{2}\to K^{0}\bar{K}^{0}]}$                 & $<4.82\times10^{-4}_{[f'_{2}\to K^{0}\bar{K}^{0}]}$                 \\ \hline
		\end{tabular}}\label{Tab:Brchi12PPP}
\end{sidewaystable}

\begin{sidewaystable}[htbp]
	\caption{Branching ratios of the  $\chi _{c2} \to P_{1}P_{2}P_{3}$ decays with the vector and tensor resonances. NWA represents the results under the narrow width approximation, and FWE represents the results under the finite width effect.}
	\centering
	\renewcommand\arraystretch{0.95}\tabcolsep 0.03in
	\resizebox{\textwidth}{!}{
		\begin{tabular}{c|l|l l|l l}
			\hline\hline
			                 Decay modes                  & Exp.data  \cite{ParticleDataGroup:2024cfk} & $V$ Resonance (NWA)                                            & $V$ Resonance (FWE)                                            & $T$ Resonance (NWA)                                                 & $T$ Resonance (FWE)                                                 \\ \hline
			                                              &                                            & $<2.96\times10^{-8}_{[\rho^{\pm}\to \pi^{\pm}\eta]}$           & $<2.62\times10^{-8}_{[\rho^{\pm}\to \pi^{\pm}\eta]}$           & $(1.34\pm0.54)\times10^{-4}_{[a^{\pm}_{2} \to \pi^{\pm}\eta]}$      & $(0.98\pm0.44)\times10^{-4}_{[a^{\pm}_{2} \to \pi^{\pm}\eta]}$      \\
			    $\chi _{c2} \to \pi^{+ }\pi^{- }\eta$     & $(4.90\pm1.30)\times10^{-4}$               & $(9.53\pm6.23)\times10^{-9}_{[\phi\to \pi^{+}\pi^{-}]}$        & $(6.70\pm4.36)\times10^{-9}_{[\phi\to \pi^{+}\pi^{-}]}$        & $(4.42\pm3.59)\times10^{-4}_{[f_{2} \to \pi^{+}\pi^{-}]}$           & $(3.17\pm2.57)\times10^{-4}_{[f_{2} \to \pi^{+}\pi^{-}]}$           \\
			                                              &                                            & $<2.96\times10^{-7}_{[\omega\to \pi^{+}\pi^{-}]}$              & $<2.08\times10^{-7}_{[\omega\to \pi^{+}\pi^{-}]}$              & $(2.79\pm2.39)\times10^{-6}_{[f'_{2} \to \pi^{+}\pi^{-}]}$          & $(1.97\pm1.69)\times10^{-6}_{[f'_{2} \to \pi^{+}\pi^{-}]}$          \\ \hline
			                                              &                                            & $(1.70\pm1.47)\times10^{-8}_{[\phi\to \pi^{+}\pi^{-}]}$        & $(1.20\pm1.04)\times10^{-8}_{[\phi\to \pi^{+}\pi^{-}]}$        & $(5.22\pm2.43)\times10^{-6}_{[a^{\pm}_{2} \to \pi^{\pm}\eta']}$     & $(4.14\pm2.06)\times10^{-6}_{[a^{\pm}_{2} \to \pi^{\pm}\eta']}$     \\
			   $\chi _{c2} \to \pi^{+ }\pi^{- }\eta'$     & $(5.10\pm1.90)\times10^{-4}$               & $<1.02\times10^{-6}_{[\omega\to \pi^{+}\pi^{-}]}$              & $<0.72\times10^{-6}_{[\omega\to \pi^{+}\pi^{-}]}$      & $<1.84\times10^{-3}_{[f_{2} \to \pi^{+}\pi^{-}]}$           & $<1.31\times10^{-3}_{[f_{2} \to \pi^{+}\pi^{-}]}$                   \\
			                                              &                                            &                                                                &                                                                & $<1.04\times10^{-5}_{[f'_{2} \to \pi^{+}\pi^{-}]}$                  & $<7.36\times10^{-6}_{[f'_{2} \to \pi^{+}\pi^{-}]}$                  \\ \hline
			  $\chi _{c2} \to \pi^{+ }\pi^{- }\pi^{0}$    & $\cdots$                                   & $(2.98\pm2.00)\times10^{-6}_{[\rho^{\pm}\to\pi^{\pm}\pi^{0}]}$ & $(2.11\pm1.41)\times10^{-6}_{[\rho^{\pm}\to\pi^{\pm}\pi^{0}]}$ & $\cdots$                                                            & $\cdots$                                                            \\
			                                              &                                            & $(2.99\pm2.00)\times10^{-6}_{[\rho^{0}\to\pi^{+}\pi^{-}]}$     & $(2.10\pm1.41)\times10^{-6}_{[\rho^{0}\to\pi^{+}\pi^{-}]}$     &                                                                     &                                                                     \\ \hline
			                                              &                                            & $<2.98\times10^{-8}_{[\rho^{0}\to \pi^{0}\eta]}$               & $<2.54\times10^{-8}_{[\rho^{0}\to \pi^{0}\eta]}$               & $(1.94\pm0.66)\times10^{-4}_{[a^{0}_{2}\to \pi^{0}\eta]}$           & $(1.41\pm0.55)\times10^{-4}_{[a^{0}_{2}\to \pi^{0}\eta]}$           \\
			     $\chi _{c2} \to \pi^{0}\pi^{0}\eta$      & $\cdots$                                   & $(4.82\pm3.15)\times10^{-9}_{[\phi\to \pi^{0}\pi^{0}]}$        & $(3.39\pm2.21)\times10^{-9}_{[\phi\to \pi^{0}\pi^{0}]}$        & $(2.23\pm1.81)\times10^{-4}_{[f_{2}\to \pi^{0}\pi^{0}]}$            & $(1.60\pm1.30)\times10^{-4}_{[f_{2}\to \pi^{0}\pi^{0}]}$            \\
			                                              &                                            & $<3.92\times10^{-9}_{[\omega\to \pi^{0}\pi^{0}]}$              & $<2.76\times10^{-9}_{[\omega\to \pi^{0}\pi^{0}]}$              & $(1.41\pm1.20)\times10^{-6}_{[f'_{2}\to \pi^{0}\pi^{0}]}$           & $(0.99\pm0.85)\times10^{-6}_{[f'_{2}\to \pi^{0}\pi^{0}]}$           \\ \hline
			                                              &                                            & $(8.55\pm7.38)\times10^{-9}_{[\phi\to \pi^{0}\pi^{0}]}$        & $(6.05\pm5.23)\times10^{-9}_{[\phi\to \pi^{0}\pi^{0}]}$        & $(7.52\pm3.07)\times10^{-6}_{[a^{0}_{2}\to \pi^{0}\eta']}$          & $(6.10\pm2.71)\times10^{-6}_{[a^{0}_{2}\to \pi^{0}\eta']}$          \\
			    $\chi _{c2} \to \pi^{0}\pi^{0}\eta'$      & $\cdots$                                   & $<1.36\times10^{-8}_{[\omega\to \pi^{0}\pi^{0}]}$              & $<1.01\times10^{-8}_{[\omega\to \pi^{0}\pi^{0}]}$              & $<9.27\times10^{-4}_{[f_{2}\to \pi^{0}\pi^{0}]}$            & $<6.60\times10^{-4}_{[f_{2}\to \pi^{0}\pi^{0}]}$                    \\
			                                              &                                            &                                                                &                                                                & $<5.22\times10^{-6}_{[f'_{2}\to \pi^{0}\pi^{0}]}$                   & $<3.70\times10^{-6}_{[f'_{2}\to \pi^{0}\pi^{0}]}$                   \\ \hline
			     $\chi _{c2} \to K^{+}K^{-}\pi^{0}$       & $(3.10\pm0.80)\times10^{-4}$               & $(2.44\pm0.38)\times10^{-5}_{[K^{*\pm}\to K^{\pm}\pi^{0}]}$    & $(1.70\pm0.26)\times10^{-5}_{[K^{*\pm}\to K^{\pm}\pi^{0}]}$    & $(3.44\pm1.41)\times10^{-5}_{[a^{0}_{2}\to K^{+}K^{-}]}$            & $(2.59\pm1.18)\times10^{-5}_{[a^{0}_{2}\to K^{+}K^{-}]}$            \\
			                                              &                                            &                                                                &                                                                & $(1.32\pm0.19)\times10^{-5}_{[K^{*\pm}_{2}\to K^{\pm}\pi^{0}]}$     & $(0.92\pm0.12)\times10^{-5}_{[K^{*\pm}_{2}\to K^{\pm}\pi^{0}]}$     \\ \hline
			                                              &                                            & $(10.04\pm1.92)\times10^{-5}_{[K^{*+}\to K^{0}\pi^{+}+c.c.]}$  & $(7.10\pm1.22)\times10^{-5}_{[K^{*+}\to K^{0}\pi^{+}+c.c.]}$   & $(9.29\pm4.35)\times10^{-5}_{[a^{+}_{2}\to K^{+}\bar{K}^{0}+c.c.]}$ & $(6.86\pm3.46)\times10^{-5}_{[a^{+}_{2}\to K^{+}\bar{K}^{0}+c.c.]}$ \\
			$\chi _{c2} \to K^{+}\bar{K}^{0}\pi^{-}+c.c.$ & $(1.30\pm0.19)\times10^{-3}$               & $(8.90\pm1.92)\times10^{-5}_{[K^{*0}\to K^{+}\pi^{-}+c.c.]}$   & $(6.00\pm1.35)\times10^{-5}_{[K^{*0}\to K^{+}\pi^{-}+c.c.]}$   & $(4.96\pm0.63)\times10^{-5}_{[K^{*+}_{2}\to K^{0}\pi^{+}+c.c.]}$    & $(3.52\pm0.45)\times10^{-5}_{[K^{*+}_{2}\to K^{0}\pi^{+}+c.c.]}$    \\
			                                              &                                            &                                                                &                                                                & $(4.24\pm0.66)\times10^{-5}_{[K^{*0}_{2}\to K^{+}\pi^{-}+c.c.]}$    & $(3.02\pm0.47)\times10^{-5}_{[K^{*0}_{2}\to K^{+}\pi^{-}+c.c.]}$    \\ \hline
			  $\chi _{c2} \to K^{0}\bar{K^{0}}\pi^{0}$    & $\cdots$                                   & $(2.11\pm0.51)\times10^{-5}_{[K^{*0}\to K^{0}\pi^{0}+c.c]}$    & $(1.49\pm0.36)\times10^{-5}_{[K^{*0}\to K^{0}\pi^{0}+c.c.]}$   & $(3.26\pm1.34)\times10^{-5}_{[a^{0}_{2}\to K^{0}K^{0}]}$            & $(2.42\pm1.09)\times10^{-5}_{[a^{0}_{2}\to K^{0}K^{0}]}$            \\
			                                              &                                            &                                                                &                                                                & $(1.05\pm0.17)\times10^{-5}_{[K^{*0}_{2}\to K^{0}\pi^{0}+c.c.]}$    & $(0.75\pm0.12)\times10^{-5}_{[K^{*0}_{2}\to K^{0}\pi^{0}+c.c.]}$    \\ \hline
			                                              &                                            &                                                                &                                                                & $(2.18\pm1.82)\times10^{-6}_{[K^{*\pm}_{2}\to K^{\pm}\eta]}$        & $(1.57\pm1.31)\times10^{-6}_{[K^{*\pm}_{2}\to K^{\pm}\eta]}$        \\
			      $\chi _{c2} \to \eta K^{+}K^{-}$        & $<3.3\times10^{-4}$                        & $(4.56\pm2.43)\times10^{-5}_{[\phi\to K^{+}K^{-}]}$            & $(3.22\pm1.72)\times10^{-5}_{[\phi\to K^{+}K^{-}]}$            & $(1.95\pm1.62)\times10^{-5}_{[f_{2}\to K^{+}K^{-}]}$                & $(1.55\pm1.28)\times10^{-5}_{[f_{2}\to K^{+}K^{-}]}$                \\
			                                              &                                            &                                                                &                                                                & $(2.07\pm1.67)\times10^{-4}_{[f'_{2}\to K^{+}K^{-}]}$               & $(1.46\pm1.18)\times10^{-4}_{[f'_{2}\to K^{+}K^{-}]}$               \\ \hline
			   $\chi _{c2} \to \eta K^{0}\bar{K^{0}}$     & $\cdots$                                   & $(2.99\pm1.58)\times10^{-5}_{[\phi\to K^{0}K^{0}]}$            & $(2.11\pm1.12)\times10^{-5}_{[\phi\to K^{0}K^{0}]}$            & $(1.83\pm1.52)\times10^{-5}_{[f_{2}\to K^{0}\bar{K}^{0}]}$          & $(1.48\pm1.23)\times10^{-5}_{[f_{2}\to K^{0}\bar{K}^{0}]}$          \\
			                                              &                                            &                                                                &                                                                & $(2.01\pm1.62)\times10^{-4}_{[f'_{2}\to K^{0}\bar{K}^{0}]}$         & $(1.42\pm1.15)\times10^{-4}_{[f'_{2}\to K^{0}\bar{K}^{0}]}$         \\ \hline
			      $\chi _{c2} \to \eta' K^{+}K^{-}$       & $(1.94\pm0.34)\times10^{-4}$               & $(7.79\pm6.30)\times10^{-5}_{[\phi\to K^{+}K^{-}]}$            & $(5.50\pm4.45)\times10^{-5}_{[\phi\to K^{+}K^{-}]}$            & $<8.16\times10^{-5}_{[f_{2}\to K^{+}K^{-}]}$                        & $<6.48\times10^{-5}_{[f_{2}\to K^{+}K^{-}]}$                        \\
			                                              &                                            &                                                                &                                                                & $<7.48\times10^{-4}_{[f'_{2}\to K^{+}K^{-}]}$                       & $<5.28\times10^{-4}_{[f'_{2}\to K^{+}K^{-}]}$                       \\ \hline
			   $\chi _{c2} \to \eta' K^{0}\bar{K^{0}}$    & $\cdots$                                   & $(5.10\pm4.11)\times10^{-5}_{[\phi\to K^{0}K^{0}]}$            & $(3.60\pm2.90)\times10^{-5}_{[\phi\to K^{0}K^{0}]}$            & $<7.68\times10^{-5}_{[f_{2}\to K^{0}\bar{K}^{0}]}$                  & $<6.14\times10^{-5}_{[f_{2}\to K^{0}\bar{K}^{0}]}$                  \\
			                                              &                                            &                                                                &                                                                & $<7.26\times10^{-4}_{[f'_{2}\to K^{0}\bar{K}^{0}]}$                 & $<5.14\times10^{-4}_{[f'_{2}\to K^{0}\bar{K}^{0}]}$                 \\ \hline
		\end{tabular}}\label{Tab:Brchi22PPP}
\end{sidewaystable}

Before discussing the numerical results, it should be emphasized that only a few resonance three-body decays of $\chi _{cJ} \to P_{1}P_{2}P_{3}$ have been measured to date. Specifically, clear resonance signals have only been reported for the $\chi _{c1} \to \pi^{+ }\pi^{-}\eta$ decay. However, the total branching ratios for many three-body $\chi_{cJ}$ decays   have been measured experimentally, rendering the exploration of potential resonance channels highly valuable and forming the primary focus of this work.

Several important physical insights can be drawn from our theoretical predictions.  First, with finite width effects included, the  branching ratios exhibit a universal reduction of approximately  20\%-30\% relative to the results obtained under the narrow width approximation. This  confirms that finite width effects are non-negligible for the three-body $\chi_{cJ}$ decay processes investigated in this work.  Second, the predicted branching ratios for the $V/T$ resonances are generally small or associated with large uncertainties. The relatively small contributions of most resonance states may explain the absence of corresponding experimental signals to date. The substantial uncertainties primarily originate from the large errors in the measured branching ratios of $\chi _{cJ}\to  PV/PT$ decays, which present theoretical challenges and limit the precision of our predictions.
Many resonance state processes have been observed in the $\chi _{c1} \to \pi^{+ }\pi^{- }\eta$ decay, and the most prominent of which is the  $a^{+}_{0}(980)$   signal, with a measured branching ratio of $(3.2\pm 0.4)\times10^{-3}$ \cite{BESIII:2016tqo}. This accounts for about 70\% of the total measured branching ratio. However, due to the lack of direct experimental measurements for the process $\chi_{cJ} \to PS$, it is difficult to make phenomenological predictions for scalar meson resonance in the absence of $\chi_{cJ} \to PS$ decays. Therefore, based on Eq.~(\ref{eq:NWA}), we extract the decay amplitude of $\chi_{c1} \to PS$ from the experimental results of cascade process  $\chi _{c1} \to a^{+}_{0}(980)\pi^{- } \to \pi^{+ }\pi^{- }\eta$, and predict the resonance contributions of other $S$ mesons. It is obvious that this approach is difficult to obtain sufficient constraints on the decay amplitudes, resulting in large uncertainty in the predicted branching ratios. The relevant results are shown in the third and fourth columns of Table \ref{Tab:Brchi12PPP}.

All scalar resonance predictions are on the order of $10^{-5}$ to $10^{-3}$. In addition to the contribution of $\chi _{c1} \to \pi^{+ }\pi^{- }\eta$ detected in the experiment, there are also some potentially valuable scalar resonance signals in other decay channels. For example,  the $f_{0}(980)$ resonance in $\chi _{c1} \to \pi^{+ }\pi^{- }\eta'$ and the $a_{0}(980)$ resonance in $K\bar{K}\pi$ channels contribute more than 80\% of the total branching ratio in their respective channels. However, experimental measurements of scalar meson resonance contributions are lacking in the $\chi_{cJ}$ decays, so the scalar resonance results  need more experimental verification. Meanwhile, we also investigate the  vector and tensor meson resonance contributions.  From the branching ratio of $\chi _{c1} \to \pi^{+ }\pi^{- }\eta$ decay, our predictions of $(1.72\pm1.31)\times10^{-4}_{[a^{\pm}_{2} \to \pi^{\pm}\eta]}$ and  $(2.72\pm0.50)\times10^{-4}_{[f_{2}\to \pi^{+}\pi^{-}]}$ under the finite width effect are very close to the experimental values of $(1.76\pm0.24)\times10^{-4}_{[a^{\pm}_{2} \to \pi^{\pm}\eta]}$ and $(3.5\pm0.6)\times10^{-4}_{[f_{2}\to \pi^{+}\pi^{-}]}$. This agreement supports the feasibility of our theoretical approach, and motivate us to identify and  analyze some potentially discovered processes based on these results.

Based on the orders of the magnitudes and the ranges of the predictions, several channels are worth further attention. The  non-resonance branching ratio of $\chi _{c2} \to \pi^{+ }\pi^{- }\pi^{0}$ has been measured with  $\mathcal{B}(\chi _{c2} \to \pi^{+ }\pi^{- }\pi^{0})^{Exp.}=(2.0\pm 0.4)\times10^{-5}$  \cite{BESIII:2016dda}. Our predicted resonance contributions are on the order of $10^{-6}$, which is one order of magnitude smaller than the non-resonance results. However, for the $\chi _{c1} \to \pi^{+ }\pi^{- }\pi^{0}$ decay, our prediction ranges are from $10^{-3}$ to $10^{-5}$, suggesting a higher likelihood of experimental detection. For the $\chi _{c1,2} \to K\bar{K}\pi$ channels, our predicted branching ratios for all resonances are $\mathcal{O}(10^{-6}-10^{-4})$.  The uncertainties in these $K\bar{K}\pi$ channels are substantially reduced due to the availability of experimental results from the corresponding two-body decay processes, rendering these decay channels highly promising for precision resonance searches at current and upcoming experimental facilities.
For the $\chi _{c1,2} \to \eta K\bar{K}$ and $\eta' K\bar{K}$ decays, the magnitudes of the theoretical central values are consistent with the experimental measurements, but the theoretical uncertainties are sizable, and they  lead to two extreme scenarios: (1) the resonance signals of $V$ and $T$ dominate (over 80\% of the total branching ratio), or (2) their resonance contributions are very small or even almost non-existent. In fact, the current phenomenological results in this work cannot fully resolve this ambiguity, which requires further relevant theoretical and experimental efforts.

In summary, the results presented in Table \ref{Tab:Brchi12PPP} and Table \ref{Tab:Brchi22PPP} demonstrate a rich spectrum of potential contributions from $S$, $V$ and $T$ resonances. These predictions offer a useful prospective reference for investigations into the $\chi _{c1,2}$ three-body decays. As more precise theoretical calculations and experimental measurements become available, the framework employed in this work can be systematically  refined to yield more accurate and valuable predictions.

\section{Conclusion} \label{sec:Conclusion}

Compared to two-body decays, the calculations of three-body or multi-body decays are more complex but offers richer dynamical information and theoretical possibilities. In the absence of first-principles calculations and  experimental data, the SU(3) flavor symmetry provides a valuable framework for connecting different decay modes and extracting insights into the underlying dynamics mechanisms. In this work, we have presented a study of three-body decays  $\chi_{c1,2} \to (V/T)P_{3}\to P_{1}P_{2}P_{3}$ and $\chi_{c1} \to SP_{3}\to P_{1}P_{2}P_{3}$ based on the SU(3) flavor symmetry/breaking approach. Some potential resonance contributions will be exhibited in the predicted results.

The predicted resonance contributions from vector and tensor mesons can be broadly categorized into two parts. One set constitutes a substantial fraction of the total branching ratio, with typically magnitudes of the order $\mathcal{O}(10^{-5}-10^{-4})$, while the other corresponds to very minor contributions, below $\mathcal{O}(10^{-7})$. Notably, the results of $\mathcal{B}(\chi _{c1,2} \to K^{+}K^{-}\pi^{0})$, $\mathcal{B}(\chi _{c1,2} \to K^{+}\bar{K}^{0}\pi^{-}+c.c.)$, and $\mathcal{B}(\chi _{c1,2} \to K^{0}\bar{K^{0}}\pi^{0})$ show a high likelihood of observable resonance signals. In most of these decay channels, the contributions from the tensor resonances dominate over those from the vector resonances. Furthermore, unlike the narrow width approximation, the finite width effects reduce the  predicted branching ratios and shrink relative uncertainties.

For the decays of charmonia $\chi_{cJ}$, where perturbative calculations face limitations, phenomenological tools like the SU(3) flavor symmetry approach offer a practical means to generate testable theoretical predictions. The results presented in this work can serve as a useful reference and are expected to be tested with future high-precision data from the BESIII experiment and the planned Super Tau-Charm Facility.

\section*{ACKNOWLEDGEMENTS}
The work was supported by the National Natural Science Foundation of China (Nos. 12365014 and 12305100).


\section*{References}


\begin{thebibliography}{99}	
\renewcommand{\baselinestretch}{1.5}
\bibitem{Mo:2006cy}
X.~H.~Mo, C.~Z.~Yuan, P.~Wang,
HEPNP \textbf{31}, 686 (2007).
arXiv:hep-ph/0611214 [hep-ph].

\bibitem{Voloshin:2007dx}
M.~B.~Voloshin,
Prog. Part. Nucl. Phys. \textbf{61}, 455 (2008).
doi:10.1016/j.ppnp.2008.02.001, arXiv:0711.4556 [hep-ph].

\bibitem{Vairo:2003gh}
A.~Vairo,
Mod. Phys. Lett. A \textbf{19}, 253 (2004).
doi:10.1142/S0217732304012927, arXiv:hep-ph/0311303 [hep-ph].

\bibitem{Novikov:1976tn}
V.~A.~Novikov, L.~B.~Okun, M.~A.~Shifman, A.~I.~Vainshtein, M.~B.~Voloshin, V.~I.~Zakharov,
Phys. Rev. Lett. \textbf{38}, 626 (1977), [erratum: Phys. Rev. Lett. \textbf{38}, 791 (1977)].

\bibitem{Reinders:1981si}
L.~J.~Reinders, H.~R.~Rubinstein, S.~Yazaki,
Nucl. Phys. B \textbf{186}, 109 (1981).

\bibitem{Nielsen:2009uh}
M.~Nielsen, F.~S.~Navarra, S.~H.~Lee,
Phys. Rept. \textbf{497}, 41 (2010).
arXiv:0911.1958 [hep-ph].

\bibitem{Wang:2012gj}
Z.~G.~Wang,
Eur. Phys. J. C \textbf{73}, 2533 (2013).
doi:10.1140/epjc/s10052-013-2533-4, arXiv:1202.2173 [hep-ph].

\bibitem{Albuquerque:2018jkn}
R.~M.~Albuquerque, J.~M.~Dias, K.~P.~Khemchandani, A.~Mart{\'\i}nez Torres, F.~S.~Navarra, M.~Nielsen, C.~M.~Zanetti,
J. Phys. G \textbf{46}, 093002 (2019).
doi:10.1088/1361-6471/ab2678, arXiv:1812.08207 [hep-ph].

\bibitem{Brambilla:2019esw}
N.~Brambilla, S.~Eidelman, C.~Hanhart, A.~Nefediev, C.~P.~Shen, C.~E.~Thomas, A.~Vairo, C.~Z.~Yuan,
Phys. Rept. \textbf{873}, 1 (2020).
doi:10.1016/j.physrep.2020.05.001, arXiv:1907.07583 [hep-ex].

\bibitem{Wu:2025hlf}
B.~Wu, Z.~Tang, R.~Rapp,
JHEP \textbf{07}, 162 (2025).
doi:10.1007/JHEP07(2025)162, arXiv:2503.10089 [nucl-th].

\bibitem{Berwein:2015vca}
M.~Berwein, N.~Brambilla, J.~Tarr{\'u}s Castell{\`a}, A.~Vairo,
Phys. Rev. D \textbf{92}, 114019 (2015).
doi:10.1103/PhysRevD.92.114019, arXiv:1510.04299 [hep-ph].

\bibitem{Liang:2016hmr}
W.~H.~Liang, J.~J.~Xie, E.~Oset,
Eur. Phys. J. C \textbf{76}, 700 (2016).
doi:10.1140/epjc/s10052-016-4563-1, arXiv:1609.03864 [hep-ph].

\bibitem{Li:2024uwu}
H.~P.~Li, J.~X.~Lin, W.~H.~Liang, E.~Oset,
Eur. Phys. J. A \textbf{61}, 97 (2025).
doi:10.1140/epja/s10050-025-01567-9, arXiv:2409.16696 [hep-ph].


\bibitem{Ablikim:2012pj}
M.~Ablikim \textit{et al.} [BESIII Collaboration],
Chin. Phys. C \textbf{37}, 063001 (2013).
doi:10.1088/1674-1137/37/6/063001, arXiv:1209.6199 [hep-ex].

\bibitem{BESIII:2017tvm}
M.~Ablikim \textit{et al.} [BESIII Collaboration],
Chin. Phys. C \textbf{42}, 023001 (2018).
doi:10.1088/1674-1137/42/2/023001, arXiv:1709.03653 [hep-ex].

\bibitem{BESIII:2024lks}
M.~Ablikim \textit{et al.} [BESIII Collaboration],
Chin. Phys. C \textbf{48}, 093001 (2024).
doi:10.1088/1674-1137/ad595b, arXiv:2403.06766 [hep-ex].

\bibitem{BES:2005ukn}
M.~Ablikim \textit{et al.} [BES Collaboration],
Phys. Lett. B \textbf{630}, 7 (2005).
doi:10.1016/j.physletb.2005.08.133, arXiv:hep-ex/0506045 [hep-ex].

\bibitem{Luchinsky:2005bf}
A.~V.~Luchinsky,
Phys. Atom. Nucl. \textbf{70}, 53 (2007).
doi:10.1134/S1063778807010073, arXiv:hep-ph/0506293 [hep-ph].

\bibitem{CLEO:2007rbf}
G.~S.~Adams \textit{et al.} [CLEO Collaboration],
Phys. Rev. D \textbf{75}, 071101 (2007).
doi:10.1103/PhysRevD.75.071101, arXiv:hep-ex/0611013 [hep-ex].

\bibitem{CLEO:2008jma}
D.~M.~Asner \textit{et al.} [CLEO Collaboration],
Phys. Rev. D \textbf{79}, 072007 (2009).
doi:10.1103/PhysRevD.79.072007, arXiv:0811.0586 [hep-ex].

\bibitem{BESIII:2010ank}
M.~Ablikim \textit{et al.} [BESIII Collaboration],
Phys. Rev. D \textbf{81}, 052005 (2010).
doi:10.1103/PhysRevD.81.052005, arXiv:1001.5360 [hep-ex].

\bibitem{Chen:2013gka}
H.~Chen, R.~G.~Ping,
Phys. Rev. D \textbf{88}, 034025 (2013).

\bibitem{Gross:2022hyw}
F.~Gross \textit{et al.},
Eur. Phys. J. C \textbf{83}, 1125 (2023).
doi:10.1140/epjc/s10052-023-11949-2, arXiv:2212.11107 [hep-ph].

\bibitem{BESIII:2025wxa}
M.~Ablikim \textit{et al.} [BESIII Collaboration],
arXiv:2512.14369 [hep-ex].

\bibitem{BESIII:2016tqo}
M.~Ablikim \textit{et al.} [BESIII Collaboration],
Phys. Rev. D \textbf{95}, 032002 (2017).
doi:10.1103/PhysRevD.95.032002, arXiv:1610.02479 [hep-ex].

\bibitem{Mo:2024zsa}
X.~H.~Mo,
Phys. Lett. B \textbf{861}, 139287 (2025).
doi:10.1016/j.physletb.2025.139287, arXiv:2501.00670 [hep-ph].

\bibitem{Haber:1985cv}
H.~E.~Haber, J.~Perrier,
Phys. Rev. D \textbf{32}, 2961 (1985).

\bibitem{Capstick:1993kb}
S.~Capstick, W.~Roberts,
Phys. Rev. D \textbf{49}, 4570 (1994).
doi:10.1103/PhysRevD.49.4570, arXiv:nucl-th/9310030 [nucl-th].

\bibitem{Wang:2016rlo}
W.~F.~Wang, H.~n.~Li,
Phys. Lett. B \textbf{763}, 29 (2016).
doi:10.1016/j.physletb.2016.10.026, arXiv:1609.04614 [hep-ph].

\bibitem{Cheng:2020iwk}
H.~Y.~Cheng, C.~W.~Chiang, C.~K.~Chua,
Phys. Rev. D \textbf{103}, 036017 (2021).
doi:10.1103/PhysRevD.103.036017, arXiv:2011.07468 [hep-ph].

\bibitem{Qi:2018lxy}
J.~J.~Qi, Z.~Y.~Wang, X.~H.~Guo, X.~W.~Kang, Z.~H.~Zhang,
Nucl. Phys. B \textbf{948}, 114788 (2019).
doi:10.1016/j.nuclphysb.2019.114788, arXiv:1811.10333 [hep-ph].

\bibitem{Lan:2025wkf}
B.~Lan, Q.~Z.~Song, J.~H.~Sheng, Y.~Qiao, R.~M.~Wang,
Chin. Phys. \textbf{49}, 123106 (2025).
doi:10.1088/1674-1137/adff00, arXiv:2508.16372 [hep-ph].

\bibitem{Wang:2012qa}
Q.~Wang, G.~Li, Q.~Zhao,
Int. J. Mod. Phys. A \textbf{27}, 1250135 (2012).
doi:10.1142/S0217751X12501357, arXiv:1202.5088 [hep-ph].

\bibitem{Cheng:2020ipp}
H.~Y.~Cheng, C.~K.~Chua,
Phys. Rev. D \textbf{102}, 053006 (2020).
doi:10.1103/PhysRevD.102.053006, arXiv:2007.02558 [hep-ph].

\bibitem{Cheng:2022ysn}
H.~Y.~Cheng,
Phys. Rev. D \textbf{106}, 113004 (2022).
doi:10.1103/PhysRevD.106.113004, arXiv:2211.03965 [hep-ph].

\bibitem{Gronau:1995hm}
M.~Gronau, O.~F.~Hernandez, D.~London, J.~L.~Rosner,
Phys. Rev. D \textbf{52}, 6356 (1995).
doi:10.1103/PhysRevD.52.6356, arXiv:hep-ph/9504326 [hep-ph].

\bibitem{Xu:2013dta}
D.~Xu, G.~N.~Li, X.~G.~He,
Int. J. Mod. Phys. A \textbf{29}, 1450011 (2014).
doi:10.1142/S0217751X14500110, arXiv:1307.7186 [hep-ph].

\bibitem{He:2014xha}
X.~G.~He, G.~N.~Li, D.~Xu,
Phys. Rev. D \textbf{91}, 014029 (2015).
doi:10.1103/PhysRevD.91.014029, arXiv:1410.0476 [hep-ph].

\bibitem{Mo:2024jjd}
X.~H.~Mo,
Phys. Rev. D \textbf{109}, 036036 (2024).
doi:10.1103/PhysRevD.109.036036, arXiv:2401.01381 [hep-ph].

\bibitem{ParticleDataGroup:2024cfk}
S.~Navas \textit{et al.} [Particle Data Group],
Phys. Rev. D \textbf{110}, 030001 (2024).
doi:10.1103/PhysRevD.110.030001.


\bibitem{Wang:2022fbk}
R.~M.~Wang, Y.~Qiao, Y.~J.~Zhang, X.~D.~Cheng, Y.~G.~Xu,
Phys. Rev. D \textbf{107}, 056022 (2023).
doi:10.1103/PhysRevD.107.056022, arXiv:2301.00090 [hep-ph].

\bibitem{BESIII:2016dda}
M.~Ablikim \textit{et al.} [BESIII Collaboration],
Phys. Rev. D \textbf{96}, 111102 (2017).
doi:10.1103/PhysRevD.96.111102, arXiv:1612.07398 [hep-ex].

\end{thebibliography}
\end{document}